\documentclass[12pt,eqno,epsf]{article}
\usepackage{amsmath,amssymb,graphicx,slashbox}
\numberwithin{equation}{section}

\def\ignore#1{{}}

\newcounter{sxn}

\newcounter{axn}

\date{}

\newdimen\mybaselineskip
\renewcommand{\thefootnote}{\arabic{footnote}}

\newcommand{\beeq}{\begin{equation}}
\newcommand{\eneq}{\end{equation}}
\newcommand{\beqn}{\begin{eqnarray}}
\newcommand{\eeqn}{\end{eqnarray}}

\newcommand{\alp}{\alpha}
\newcommand{\bt}{\beta}

\newcommand{\dlt}{\delta}

\newcommand{\ep}{\epsilon}
\newcommand{\tht}{\theta}

\newcommand{\vth}{\vartheta}

\newcommand{\lmd}{\lambda}
\newcommand{\Lmd}{\Lambda}
\newcommand{\sgm}{\sigma}
\newcommand{\Sgm}{\Sigma}
\newcommand{\Ups}{\Upsilon}
\newcommand{\vph}{\varphi}

\newcommand{\Omg}{\Omega}
\newcommand{\dalp}{\dot{\alpha}}
\newcommand{\dbt}{\dot{\beta}}

\newcommand{\be}{\begin{equation}}
\newcommand{\ee}{\end{equation}}
\newcommand{\bea}{\begin{eqnarray}}
\newcommand{\eea}{\end{eqnarray}}
\newcommand{\eql}{\!\!\!&=\!\!\!&}

\newcommand{\defa}{\!\!\!&\equiv\!\!\!&}

\newcommand{\toa}{\!\!\!&\to\!\!\!&}

\newcommand{\exch}{\leftrightarrow}

\newcommand{\tl}[1]{\tilde{#1}}
\newcommand{\bdm}[1]{{\mbox{\boldmath $#1$}}}

\newcommand{\diag}{{\rm diag}}
\newcommand{\der}{\partial}
\newcommand{\dr}{\!\!d}
\newcommand{\hc}{{\rm h.c.}}
\newcommand{\ie}{{i.e.}}
\newcommand{\id}{\mbox{\boldmath $1$}}

\newcommand{\udl}[1]{\underline{#1}}

\newcommand{\vev}[1]{\langle #1 \rangle}
\newcommand{\Lvev}[1]{\left\langle #1 \right\rangle}
\newcommand{\brkt}[1]{\left( #1 \right)}
\newcommand{\brc}[1]{\left\{ #1 \right\}}
\newcommand{\sbk}[1]{\left[ #1 \right]}
\newcommand{\abs}[1]{\left| #1 \right|}

\renewcommand{\Re}{{\rm Re}\,}
\renewcommand{\Im}{{\rm Im}\,}

\newcommand{\cC}{{\cal C}}

\newcommand{\cE}{{\cal E}}

\newcommand{\cH}{{\cal H}}

\newcommand{\cL}{{\cal L}}

\newcommand{\cN}{{\cal N}}
\newcommand{\cO}{{\cal O}}
\newcommand{\cP}{{\cal P}}
\newcommand{\cQ}{{\cal Q}}
\newcommand{\cR}{{\cal R}}

\newcommand{\cV}{{\cal V}}

\newcommand{\cW}{{\cal W}}
\newcommand{\cX}{{\cal X}}
\newcommand{\cY}{{\cal Y}}
 
\newcommand{\bE}{{\mathbb E}}

\newcommand{\bH}{{\mathbb H}}

\newcommand{\bT}{{\mathbb T}}
\newcommand{\bV}{{\mathbb V}}
\newcommand{\rmE}{{\rm E}}
\newcommand{\rmT}{{\rm T}}
\newcommand{\SE}{S_{\rmE}}
\newcommand{\VE}{V_{\rmE}}
\newcommand{\derE}{\der_{\rmE}}
\newcommand{\UE}{R_{\rm E}}

\newcommand{\suU}{\mbox{SU(2)}_{\mbox{\scriptsize\bf U}}}

\begin{document}
\thispagestyle{empty}

\baselineskip=12pt


\begin{flushright}
KEK-TH-1991 \\
WU-HEP-17-13
\end{flushright}

\baselineskip=35pt plus 1pt minus 1pt

\vskip 1.5cm

\begin{center}
{\LARGE\bf Full diffeomorphism and Lorentz invariance
in 4D $\bdm{\cN=1}$ superfield description \\
of 6D SUGRA}

\vspace{1.5cm}
\baselineskip=20pt plus 1pt minus 1pt

\normalsize

{\large\bf Hiroyuki Abe,}${}^1\!${\def\thefootnote{\fnsymbol{footnote}}
\footnote[1]{E-mail address: abe@waseda.jp}}
{\large\bf Shuntaro Aoki}${}^1\!${\def\thefootnote{\fnsymbol{footnote}}
\footnote[2]{E-mail address: shun-soccer@akane.waseda.jp}}
{\large\bf and Yutaka Sakamura}${}^{2,3}\!${\def\thefootnote{\fnsymbol{footnote}}
\footnote[3]{E-mail address: sakamura@post.kek.jp}}

\vskip 1.0em

${}^1${\small\it Department of Physics, Waseda University, \\ 
Tokyo 169-8555, Japan}

\vskip 1.0em

${}^2${\small\it KEK Theory Center, Institute of Particle and Nuclear Studies, 
KEK, \\ Tsukuba, Ibaraki 305-0801, Japan} \\ \vspace{1mm}
${}^3${\small\it Department of Particles and Nuclear Physics, \\
SOKENDAI (The Graduate University for Advanced Studies), \\
Tsukuba, Ibaraki 305-0801, Japan}

\end{center}

\vskip 1.0cm
\baselineskip=20pt plus 1pt minus 1pt

\begin{abstract}
We complete the four-dimensional $\cN=1$ superfield description of six-dimensional supergravity. 
The missing ingredients in the previous works are the superfields 
that contain the sechsbein~$e_4^{\;\;\udl{\nu}}$, $e_5^{\;\;\udl{\nu}}$, $e_\mu^{\;\;\udl{4}}$, 
$e_\mu^{\;\;\udl{5}}$ and the second gravitino. 
They are necessary to make the action invariant under 
the diffeomorphisms and the Lorentz transformations involving the extra dimensions. 
We find the corresponding superfield transformation laws, 
and show the invariance of the action under them. 
We also check that the resultant action reproduces the known superfield description 
of five-dimensional supergravity 
through the dimensional reduction. 
\end{abstract}

\newpage

\section{Introduction}
When we consider higher-dimensional supersymmetric (SUSY) theories, 
it is useful to describe the action in terms of $\cN=1$ superfields~\cite{Marcus:1983wb}-\cite{Abe:2004ar} 
for various reasons.\footnote{
``$\cN=1$'' denotes SUSY with four supercharges in this paper. 
} 
It makes the expression of the action much more compact than 
the component field expression. 
In particular, the complicated spacetime index structures become much simpler. 
In higher than six dimensions (6D), however, the full superspace formulation is not known 
due to the extended SUSY structure. 
Even in such cases, the $\cN=1$ superfield expression is still possible 
because only partial SUSY structure is respected. 
Such an expression is useful to discuss a system 
in which the spacetime is compactified to four dimensions (4D)
and the $\cN=1$ SUSY is preserved. 
We can derive the 4D effective action directly from the higher dimensional theory, 
keeping the $\cN=1$ superspace structure manifest. 
Especially, when the system contains lower dimensional branes or orbifold fixed points 
in the compactified space, 
the bulk-brane interactions are described in a transparent manner 
because all the sectors are expressed on the common $\cN=1$ superspace. 
Besides, the $\cN=1$ superfield formalism is familiar to many researchers, 
and is easy to handle. 

For global SUSY theories, the $\cN=1$ superfield description of the action has 
been already provided in 5-10 dimensions~\cite{ArkaniHamed:2001tb}. 
We have to extend it to the supergravity (SUGRA) 
in order to discuss the moduli stabilization, 
the interactions to the moduli or the higher dimensional gravitational multiplet, and so on. 
However, such an extension is not straightforward. 
First, it is a nontrivial task to identify the component fields of the $\cN=1$ superfields. 
It usually happens that the non-gravitational fields form the superfields 
with the help of the gravitational fields, such as the vierbein and the gravitini. 
Of course, these superfields should reduce to the ones in Ref.~\cite{ArkaniHamed:2001tb} 
if the gravitational fields are replaced with their background values in the flat spacetime. 
However, such an observation alone is not enough to identify the dependence of 
each component of the superfield on the gravitational fields. 
The complete identification can be achieved by requiring the invariance of the action 
under various symmetry transformations, 
such as the gauge transformations, the diffeomorphisms, the Lorentz transformations, etc. 
We should note that the diffeomorphisms and the Lorentz transformations have to 
be divided into the 4D parts and the extra-dimensional parts, and treated separately 
because we only respect the $\cN=1$ SUSY. 
The invariance under their 4D parts is obvious. 
In contrast, the invariances under the diffeomorphism in the extra dimensions 
and the Lorentz transformations that mix the 4D index with the extra-dimensional one 
are less trivial, but they are also expressed as the $\cN=1$ superfield transformations. 
Besides, we should also note that the $\cN=1$ superconformal parameters~\footnote{
4D $\cN=1$ SUGRA can be described by the superconformal 
formulation~\cite{Kaku:1978nz,Kaku:1978ea,Kugo:1982cu}, 
which is also expressed by the corresponding superspace formulation~\cite{Butter:2009cp,Sakamura:2011df}. 
} 
depend on the extra-dimensional coordinates, and that the desired superfield action involves  
the derivatives with respective to such coordinates. 
Therefore, we need to covariantize such derivatives. 
The corresponding connection superfields contain the ``off-diagonal'' components of 
the vierbein~$e_{\mu}^{\;\;\udl{n}}$ and $e_m^{\;\;\udl{\nu}}$, 
where $\{\mu,\nu\}$ and $\{m,n\}$ denote the 4D and the extra-dimensional indices, 
respectively. 

The simplest background for the extra-dimensional models is the five-dimensional (5D) spacetime. 
The $\cN=1$ description of the 5D SUGRA action is provided in Refs.~\cite{Paccetti:2004ri,Abe:2004ar}. 
These works specify the dependence of the action on the ``modulus'' superfields 
that contains the extra-dimensional component of the f\"unfbein~$e_4^{\;\;\udl{4}}$. 
This superfield description makes it possible to derive 
the 4D effective action for various setups 
systematically~\cite{Abe:2006eg,Correia:2006pj,Abe:2008an,Abe:2011rg}. 
However, the superfield action in Refs.~\cite{Paccetti:2004ri,Abe:2004ar} does not contain 
the ``off-diagonal'' components of the f\"unfbein~$e_\mu^{\;\;\udl{4}}$, $e_4^{\;\;\udl{\nu}}$ 
and their $\cN=1$ SUSY partners. 
Thus, the action is not invariant under the diffeomorphism in the extra dimension and 
the Lorentz transformations that mix the 4D and the fifth dimensions. 
Those missing ingredients are incorporated at the linearized level in Ref.~\cite{Sakamura:2012bj}, 
and play an important role in the calculation of 
the one-loop effective potential~\cite{Sakamura:2013cqd,Sakamura:2013rba,Sakamura:2014aja}. 

In this paper, we focus on 6D SUGRA~\cite{Nishino:1984gk,Salam:1984cj,Bergshoeff:1985mz}. 
The 6D spacetime is the next simplest setup for the extra-dimensional models, 
and the minimal setup where the shape modulus for the extra-dimensional space appears. 
6D SUGRA generically contains the Weyl multiplet as the gravitational multiplet, 
and $n_{\rm H}$ hypermultiplets, $n_{\rm V}$ vector multiplets 
and $n_{\rm T}$ tensor multiplets as the mattter multiplets. 
From the anomaly cancellation condition, the numbers of the multiplets are constrained by
$29n_{\rm T}+n_{\rm H}-n_{\rm V}=273$~\cite{RandjbarDaemi:1985wc,Green:1984bx,Kumar:2010ru}. 
In contrast to 5D SUGRA, the Weyl multiplet contains 
the anti-self-dual tensor field~$T_{MN}^-$ ($M,N=0,1,\cdots,5$), 
and a 6D tensor multiplet contains the self-dual tensor field~$B_{MN}^+$. 
In general, the (anti-)self-dual condition is an obstacle to the Lagrangian formulation, 
similar to that of type IIB SUGRA. 
However, when $n_{\rm T}=1$, this difficulty can be solved 
because we can construct an unconstrained tensor field~$B_{MN}$ 
by combining $T_{MN}^-$ with $B_{MN}^+$~\cite{Bergshoeff:1985mz,Coomans:2011ih}. 
When $n_{\rm T}\neq 1$, the (anti-)self-dual conditions remain, 
and thus the theory cannot be described by the Lagrangian. 
Hence, we focus on the case of $n_{\rm T}=1$ in this paper. 

In our previous work~\cite{Abe:2015bqa}, we found the $\cN=1$ superfield description 
of the vector-tensor couplings in 6D global SUSY theories, which is derived from 
the invariant action~\cite{Linch:2012zh} 
in the projective superspace~\cite{Karlhede:1984vr,Lindstrom:1987ks,Lindstrom:1989ne}.\footnote{
6D projective superspace is also discussed in Refs.~\cite{Grundberg:1984xr,Gates:2005mc,Gates:2006it}. 
} 
Then, we extend this result to 6D SUGRA in Ref.~\cite{Abe:2015yya} 
by identifying the ``moduli superfields'' 
that contain the extra-dimensional components of the sechsbein~$e_m^{\;\;\udl{n}}$ 
($m,n=4,5$), and inserting them into the result in Ref.~\cite{Abe:2015bqa}. 
We have checked that the resultant action is invariant under the supergauge transformation, 
and reproduces the known 5D SUGRA action after the dimensional reduction. 
In this paper, we complete the $\cN=1$ superfield description of 6D SUGRA 
by incorporating the missing ingredients, 
\ie, the ``off-diagonal'' components of the sechsbein~$e_\mu^{\;\;\udl{n}}$ and $e_m^{\;\;\udl{\nu}}$ 
($m,n=4,5$) and their $\cN=1$ superpartners. 
The identification of the corresponding superfields and 
the dependence of the action on them are determined 
by the invariance under the full 6D diffeomorphisms. 
These newly incorporated superfields, 
which are the real superfields~$U^m$ and the spinor superfields~$\Psi_m^\alp$ ($m=4,5$), 
are also necessary for the invariance 
under the Lorentz transformations that mix the 4D and the extra-dimensional indices. 
This work corresponds to the 6D extension of Ref.~\cite{Sakamura:2012bj}. 
We will treat the 4D $\cN=1$ SUGRA part at the linearized level for a technical reason. 
Due to this approximation, we can only determine the dependence of the action on $\Psi_m^\alp$ 
at the linearized level. 
In contrast, we clarify the dependence on $U^m$ at the full order~\footnote{
Some of the $U^m$-dependent terms are treated at the linearized level 
due to technical difficulties. 
}  
because it is determined only by the invariance 
under diffeomorphisms in the extra dimensions, independently of the 4D diffeomorphism.

The paper is organized as follows. 
We provide a brief review of our previous work~\cite{Abe:2015yya} 
in the next section. 
In Sec.~\ref{Xi-invariance}, we require the invariance of the action 
under the diffeomorphisms in the extra dimensions, 
and introduce the connection superfields~$U^m$ ($m=4,5$) that contain 
the ``off-diagonal'' components of the sechsbein. 
In Sec.~\ref{cov_derE}, we covariantize the derivatives with respective to 
the extra-dimensional coordinates by introducing another connection superfields~$\Psi_m^\alp$ ($m=4,5$). 
In Sec.~\ref{Lorentz_inv}, we address the Lorentz transformations that mix 
the 4D and the extra-dimensional indices, and show the invariance of the action under them. 
In Sec.~\ref{dim_red:5D}, we check that the resultant superfield action of 6D SUGRA 
reduces to the known 5D SUGRA action after the dimensional reduction. 
Sec.~\ref{summary} is devoted to the summary. 
In Appendix~\ref{4DSUGRAcouplings}, we collect the results of Ref~\cite{Sakamura:2011df} that discusses 
the 4D linearized SUGRA and the superfield description of the $\cN=1$ superconformal transformation. 
In Appendices~\ref{diffeo:comp} and \ref{Lorentz:comp}, 
we show the diffeomorphisms and the Lorentz transformations
in the component field expression, 
and provide the correspondence to the superfield description.

\section{Review of our previous work}
The 6D spacetime indices~$M,N,\cdots=0,1,2,\cdots,5$ are divided 
into the 4D part~$\mu,\nu,\cdots =0,1,2,3$ and the extra-dimensional part~$m,n,\cdots=4,5$. 
The corresponding local Lorentz indices are denoted by the underbarred ones. 
We assume that the 4D part of the spacetime has the flat background geometry, 
and follow the notation of Ref.~\cite{Wess:1992cp} for the 2-component spinors. 

\subsection{$\bdm{\cN=1}$ decomposition of 6D supermultiplets}
The 6D Weyl multiplet~$\bE$ consists of the sechsbein~$e_M^{\;\;\underline{N}}$, 
the gravitino~$\psi_{M\bar{\alp}}^i$, 
the $\suU$ (auxiliary) gauge fields~$V_M^{ij}$, and the other auxiliary fields, 
where $\bar{\alp}$ is a 6D spinor index, and $i,j=1,2$ are the $\suU$-doublet indices. 
The gravitino has the 6D chirality~$+$, 
and is the $\suU$-Majorana-Weyl fermion, which can be decomposed into 
the two 4D Dirac fermions. 
\be
 \psi_M^1= \begin{pmatrix} \psi_{M\alp}^+ \\ \bar{\psi}^{-\dalp}_M \end{pmatrix}, \;\;\;\;\;
 \psi_M^2 = \begin{pmatrix} -\psi_{M\alp}^- \\ \bar{\psi}^{+\dalp}_M \end{pmatrix}, 
\ee
where $\alp,\dalp=1,2$ are the 2-component spinor indices. 
If we choose $\ep^+_\alp$ and $\bar{\ep}^{+\dalp}$ in the 6D SUSY transformation 
parameter~$\ep^i_{\bar{\alp}}$ as the 4D $\cN=1$ SUSY one we respect, 
the fields~$\brc{e_\mu^{\;\;\udl{\nu}},\psi_\mu^+,\cdots}$ form the 4D Weyl multiplet. 
We can construct the real superfield~$U^\mu$ from them as 
(see Appendix~\ref{4DSUGRAcouplings}) 
\be
 U^\mu = (\tht\sgm^\nu\bar{\tht})\tl{e}_\nu^{\;\;\mu}
 +i\bar{\tht}^2\brkt{\tht\sgm^\nu\bar{\sgm}^\mu\psi_\nu^+}
 -i\tht^2\brkt{\bar{\tht}\bar{\sgm}^\nu\sgm^\mu\bar{\psi}_\nu^+}+\cdots, 
\ee
where $\tl{e}_\nu^{\;\;\mu}$ is the fluctuation field around the background 
defined as (\ref{def:tle}), and 
\be
 \sgm^\mu\equiv\vev{e_{\udl{\nu}}^{\;\;\mu}}\sgm^{\udl{\nu}}, \;\;\;\;\;
 \bar{\sgm}^\mu\equiv\vev{e_{\udl{\nu}}^{\;\;\mu}}\bar{\sgm}^{\udl{\nu}}. 
\ee
Note that we need not discriminate the flat and the curved 4D indices 
for $\tl{e}_\nu^{\;\;\mu}$ at the linearized order 
since the 4D part of the background spacetime is assumed to be flat 
($\vev{e_\nu^{\;\;\underline{\mu}}}=\dlt_\nu^{\;\;\mu}$). 
As explicitly shown in Appendix~\ref{4DInv_action}, 
once the matter action is given, we can always obtain its 4D gravitational couplings. 
Thus, we will omit the dependences on $U^\mu$ 
to simplify the expressions in the following. 

In our previous work~\cite{Abe:2015yya}, we have found that 
the extra dimensional components of the sechsbein~$e_m^{\;\;\udl{n}}$ and its superpartners form 
the chiral superfield~$\SE$ and the real general superfield~$\VE$ as
\bea
 \SE \eql \sqrt{\frac{E_4}{E_5}}
 +\cO(\tht), \nonumber\\
 \VE \eql e^{(2)}+\cO(\tht), 
\eea
where $E_m\equiv e_m^{\;\;\udl{4}}+ie_m^{\;\;\udl{5}}$ 
and $e^{(2)}\equiv \det(e_m^{\;\;\udl{n}}) = e_4^{\;\;\udl{4}}e_5^{\;\;\udl{5}}-e_4^{\;\;\udl{5}}e_5^{\;\;\udl{4}}$. 
These correspond to the shape and the volume moduli, respectively. 

The matter field content consists of hypermultiplets~$\mathbb{H}^A$ ($A=1,2,\cdots,n_{\rm H}$), 
vector multiplets $\mathbb{V}^I$ ($I=1,2,\cdots,n_{\rm V}$),\footnote{
The anomaly cancellation conditions constrain the numbers of the multiplets (see the introduction) 
and the gauge group~\cite{RandjbarDaemi:1985wc,Green:1984bx,Kumar:2010ru}. 
In this paper, we do not consider such constraints, and assume that the gauge groups are Abelian, 
for simplicity. 
} and a tensor multiplet~$\bT$. 
They are decomposed into $\cN=1$ superfields as
\be
 \bH^A = (H^{2A-1},H^{2A}), \;\;\;\;\;
 \bV^I = (V^I,\Sgm^I), \;\;\;\;\;
 \bT = (\Ups_{\rmT\alp},V_{{\rm T}4},V_{{\rm T}5}), 
\ee
where $H^{2A-1},H^{2A},\Sgm^I,\Ups_{\rmT\alp}$ are chiral superfields, 
and $V^I$, $V_{\rmT 4}$ and $V_{\rmT 5}$ are real superfields. 
Here, $\bH^A$ contains the hyperscalars~$(\phi_i^{2A-1},\phi_i^{2A})$, which is subject to
the reality condition:~$\brkt{\phi_1^{2A-1}}^*=\phi_2^{2A}$, $\brkt{\phi_1^{2A}}^*=-\phi_2^{2A-1}$, 
$\bV^I$ contains a 6D vector field~$A_M^I$, 
and $\bT$ contains a real scalar field~$\sgm$ and an anti-symmetric tensor field~$B_{MN}$. 
The hypermultiplets~$\bH^A$ are divided into the compensator multiplets~$A=1,2,\cdots,n_{\rm comp}$ 
and the physical ones~$A=n_{\rm comp}+1,\cdots,n_{\rm comp}+n_{\rm phys}$. 
The lowest bosonic components of the superfields are~\footnote{
The factor~$i/2$ was missing for the lowest component of $\Sgm^I$ in Ref.~\cite{Abe:2015yya}. 
Besides, $V_{\rmT m}=-8X_m$ ($m=4,5$) and $\Ups_{\rmT\alp}=8\bar{D}^2Y_\alp$ 
in the notation of Ref.~\cite{Abe:2015yya}. 
}
\bea
 H^{\bar{A}} \eql \brkt{E_4E_5}^{1/4}\phi_2^{\bar{A}}+\cO(\tht), \nonumber\\
 V^I \eql -(\tht\sgm^{\udl{\mu}}\bar{\tht})A_{\udl{\mu}}^I+\cO(\tht^3), \nonumber\\
 \Sgm^I \eql \frac{i}{2}\brkt{\frac{1}{\SE|}A_4^I-\SE| A_5^I}+\cO(\tht), \nonumber\\
 \Ups_{\rmT\alp} \eql -\tht_\alp\brkt{2B_{\udl{4}\udl{5}}+i\sgm}
 -2i\brkt{\sgm^{\udl{\mu}\udl{\nu}}\tht}_\alp B_{\udl{\mu}\udl{\nu}}+\cO(\tht^2), \nonumber\\
 V_{\rmT m} \eql -2(\tht\sgm^{\udl{\mu}}\bar{\tht})B_{\udl{\mu} m}+\cO(\tht^3), \;\;\;\;\;
 (m = 4,5)  
%
%
%
%
 \label{sf:components}
\eea
where $\bar{A}=2A-1,2A$, and $\SE|=\sqrt{E_4/E_5}$ is the lowest component of $S_E$. 

The supergauge transformations are given by 
\be
 \dlt_\Lmd V^I = \Lmd^I+\bar{\Lmd}^I, \;\;\;\;\;
 \dlt_\Lmd\Sgm^I = \derE\Lmd^I, 
\ee
where the transformation parameters~$\Lmd^I$ are chiral superfields, and
\be
 \derE \equiv \frac{1}{\SE}\der_4-\SE\der_5. 
\ee
The gauge-invariant field strength superfields are given by 
\be
 \cW_\alp^I \equiv -\frac{1}{4}\bar{D}^2D_\alp V^I. \label{def:cW_alp^I:1}
\ee

The SUSY extension of the tensor gauge transformation:~$B_{MN}
\to B_{MN}+\der_M\lmd_N-\der_N\lmd_M$ 
($\lmd_M$: real parameter) is expressed as
\bea
 \dlt_{\rm G}V_{\rmT 4} \eql -\der_4 V_{\rm G}+\Re(\SE\Sgm_{\rm G}), \;\;\;\;\;
 \dlt_{\rm G}V_{\rmT 5} = -\der_5 V_{\rm G}+\Re\brkt{\frac{\Sgm_{\rm G}}{\SE}}, \nonumber\\
 \dlt_{\rm G}\Ups_{\rmT\alp} \eql -\frac{1}{4}\bar{D}^2D_\alp V_{\rm G}, 
\eea
where the transformation parameters~$V_{\rm G}$ and $\Sgm_{\rm G}$ are 
a real and a chiral superfields respectively, which form a 6D vector multiplet~$\bV_{\rm G}$.  
\bea
 V_{\rm G} \eql -2(\tht\sgm^{\udl{\mu}}\bar{\tht})\lmd_{\udl{\mu}}+\cO(\tht^3), \nonumber\\
 \Sgm_{\rm G} \eql \frac{2\abs{\SE|}^2}{\Im\SE^2|}
 \brkt{\frac{1}{\bar{S}_{\rm E}|}\lmd_4-\bar{S}_{\rm E}|\lmd_5}+\cO(\tht). 
\eea
The superfields other than $\bT$ are neutral. 
The field strengths invariant under this transformation are 
\bea
 \cX_{\rmT} \defa \frac{1}{2}\Im\brkt{D^\alp\Ups_{\rmT\alp}}, \nonumber\\
 \cY_{\rmT\alp} \defa \frac{1}{2\SE}\cW_{\rmT 4\alp}+\frac{\SE}{2}\cW_{\rmT 5\alp}
 +\frac{1}{2}\SE\cO_{\rmE}\Ups_{\rmT\alp},  \label{def:cXcY}
\eea
where 
\bea
 \cW_{\rmT m\alp} \defa -\frac{1}{4}\bar{D}^2D_\alp V_{\rmT m}, \;\;\;\;\; (m=4,5) \nonumber\\
 \cO_{\rmE} \defa \frac{1}{\SE^2}\der_4+\der_5.  \label{def:cW_Tmalp}
\eea
Namely, $\cX_{\rmT}$ and $\cY_{\rmT\alp}$ are real linear and chiral superfields, respectively. 
The tensor multiplet~$(\Ups_{\rmT\alp},V_{\rmT m})$ is subject to the constraints: 
\bea
 &&\frac{1}{\SE}\cW_{\rmT 4\alp}-\SE\cW_{\rmT 5\alp}+\derE\Ups_{\rmT\alp} = 0, \nonumber\\
 &&\bar{D}^2D_\alp\brkt{\cX_{\rmT}\VE} = -4\brc{\der_E\cY_{\rmT\alp}-(\cO_{\rmE}\SE)\cY_{\rmT\alp}}. 
 \label{Tconstraints}
\eea

In the global SUSY limit, these constraints reduce to the superfield version of 
the self-dual condition: 
\be
 \der_{[M} B_{NL]}^+=\frac{1}{6}\ep_{MNLPQR}\der^P B^{+QR}.  \label{self-dual}
\ee
In fact, in the limit of $\SE\to e^{-i\pi/4}$ and $V_{\rm E}\to 1$, 
(\ref{Tconstraints}) is reduced to  
\bea
 &&\cW_{\rmT 4\alp}+i\cW_{\rmT 5\alp}+\brkt{\der_4+i\der_5}\Ups_{\rmT\alp} = 0, \nonumber\\
 &&\bar{D}^2D_\alp\cX_{\rmT} = -4e^{i\pi/4}\brkt{\der_4+i\der_5}\cY_{\rmT\alp}. 
 \label{Tconstraints:2}
\eea
The field strength superfield~$\cY_{\rmT\alp}$ becomes
\bea
 \cY_{\rm T\alp} \eql \frac{e^{i\pi/4}}{2}\brc{\cW_{\rmT 4\alp}-i\cW_{\rmT 5\alp}
 +\brkt{\der_4-i\der_5}\Ups_{\rmT\alp}} \nonumber\\
 \eql e^{i\pi/4}\brkt{\cW_{\rmT 4\alp}+\der_4\Ups_{\rmT\alp}}
 = e^{-i\pi/4}\brkt{\cW_{\rmT 5\alp}+\der_5\Ups_{\rmT\alp}}. 
\eea
In the second line, we have used the first constraint in (\ref{Tconstraints:2}). 
From these expressions, we obtain 
\be
 D^\alp\cY_{\rmT\alp} = -2e^{-i\pi/4}\brkt{\der_4-i\der_5}\cX_{\rmT}. 
\ee
This and the second constraint in (\ref{Tconstraints:2}) contain 
the self-dual condition~(\ref{self-dual}). 
Thus, the antisymmetric tensor~$B_{MN}$ in $\Ups_{\rmT\alp}$ and $V_{\rmT m}$ becomes
the self-dual tensor~$B^+_{MN}$ in the global SUSY limit.

\ignore{
The first constraint indicates that
\be
 D^\alp\brkt{\UE^2\cY_{\rmT\alp}} = -2\bar{\der}_{\rmE}\cX_{\rmT}
 +\frac{i\bar{D}_{\dalp}\bar{S}_{\rmE}}{\bar{S}_{\rmE}}\bar{\cY}_{\rmT}^{\dalp}, 
\ee
where
\be
 \UE \equiv \brkt{\Im\frac{\bar{S}_{\rmE}}{\SE}}^{1/2}. \label{def:U_E}
\ee
This and the second equation in (\ref{Tconstraints}) represent  
the SUSY extension of the self-dual condition for $B_{MN}^+$. 
}

In the SUGRA case, the second constraint in (\ref{Tconstraints}) can be solved as follows. 
Using the first constraint in (\ref{Tconstraints}), $\cY_{\rmT\alp}$ can be expressed as
\bea
 \cY_{\rmT\alp} \eql \frac{1}{\SE}\brkt{\cW_{\rmT 4\alp}+\der_4\Ups_{\rmT\alp}} \nonumber\\
 \eql \SE\brkt{\cW_{\rmT 5\alp}+\der_5\Ups_{\rmT\alp}}. 
\eea
Thus the second constraint in (\ref{Tconstraints}) is rewritten as
\bea
 \bar{D}^2D_\alp\brkt{\cX_{\rmT}\VE} \eql -4\der_4\brkt{\frac{\cY_{\rmT\alp}}{\SE}}
 +4\der_5\brkt{\SE\cY_{\rmT\alp}} \nonumber\\
 \eql -4\der_4\brkt{\cW_{\rmT 5\alp}+\der_5\Ups_{\rmT\alp}}
 +4\der_5\brkt{\cW_{\rmT 4\alp}+\der_4\Ups_{\rmT\alp}} \nonumber\\
 \eql \bar{D}^2D_\alp\brkt{\der_4 V_{\rmT 5}-\der_5 V_{\rmT 4}}, 
\eea
which can be solved as 
\be
 \cX_{\rmT}\VE = \der_4V_{\rmT 5}-\der_5V_{\rmT 4}+\Sgm_{\rmT}+\bar{\Sgm}_{\rmT} 
 \equiv \cV_{\rmT},  \label{def:cV_T}
\ee
where $\Sgm_{\rmT}$ is a chiral superfield. 
The lowest component of $\Sgm_{\rmT}$ is identified as
\be
 \Sgm_{\rmT}| = \frac{1}{2}e^{(2)}\sgm-iB_{45}. \label{comp:Sgm_T}
\ee
Eq.~(\ref{def:cV_T}) indicates that the ``volume modulus'' superfield~$\VE$ is expressed 
by $\Ups_{\rm T\alp}$, $V_{\rmT m}$ and $\Sgm_{\rmT}$, 
and is not an independent degree of freedom.

\subsection{Invariant action}
The $\cN=1$ superfield description of (the $U^\mu$-independent part of) the 6D SUGRA action
provided in Ref.~\cite{Abe:2015yya} is 
\bea
 S \eql \int\dr^6x\;\brkt{\cL_{\rm H}+\cL_{\rm VT}}, \nonumber\\
 \cL_{\rm H} \eql -\int\dr^4\tht\;2\VE^{1/2}\UE^{1/2}
 \brkt{H_{\rm odd}^\dagger\tl{d}e^VH_{\rm odd}+H_{\rm even}^\dagger\tl{d}e^{-V}H_{\rm even}} 
 \nonumber\\
 &&+\sbk{\int\dr^2\tht\;\brc{H_{\rm odd}^t\tl{d}\brkt{\derE-\Sgm}H_{\rm even}
 -H_{\rm even}^t\tl{d}\brkt{\derE+\Sgm}H_{\rm odd}}+\hc}, \nonumber\\
 \cL_{\rm VT} \eql \int\dr^4\tht\;f_{IJ}\left[
 \brc{-2\Sgm^ID^\alp V^J\cY_{\rmT\alp}+\frac{1}{2}
 \brkt{\derE V^ID^\alp V^J-\derE D^\alp V^IV^J}\cY_{\rmT\alp}+\hc} \right.\nonumber\\
 &&\hspace{20mm}
 +\cX_{\rmT}\VE\brkt{D^\alp V^I\cW_\alp^J
 +\frac{1}{2}V^ID^\alp\cW_\alp^J+\hc} \nonumber\\
 &&\hspace{20mm}
 +\frac{\cX_{\rmT}}{\UE}\left\{\rule{0pt}{15pt}
 4\brkt{\bar{\der}_{\rmE}V^I-\bar{\Sgm}^I}\brkt{\derE V^J-\Sgm^J}
 -2\bar{\der}_{\rmE}V^I\derE V^J \right. \nonumber\\
 &&\hspace{32mm}\left.\left.
 +\frac{2\SE}{\bar{S}_{\rmE}}\Sgm^I\Sgm^J
 +\frac{2\bar{S}_{\rmE}}{\SE}\bar{\Sgm}^I\bar{\Sgm}^J\right\}\right], 
 \label{prev:S}
\eea
where $H_{\rm odd}=(H^1,H^3,H^5,\cdots)^t$, $H_{\rm even}=(H^2,H^4,H^6,\cdots)^t$, 
$\tl{d}=\diag(\id_{n_{\rm comp}},-\id_{n_{\rm phys}})$ is the metric of the hyperscalar space 
that discriminates the compensator multiplets from the physical ones,\footnote{
In contrast to 4D SUGRA, an arbitrary number of the compensators is possible 
in 5D and 6D SUGRAs. 
When $n_{\rm comp}>1$, the superconformal gauge-fixing conditions cannot eliminate 
all the degrees of freedom of the compensators. 
So some auxiliary multiplets are necessary to eliminate them. 
(See Ref.~\cite{Fujita:2001bd}, for example.) 
The number~$n_{\rm comp}$ determines the geometry of 
the space spanned by the physical hyperscalars. 
}
$f_{IJ}=f_{JI}$ are real constants, and~\footnote{
$\UE$ is denoted as $U_{\rm E}^2$ in Ref.~\cite{Abe:2015yya}. 
}
\be
 \UE \equiv \Im\frac{\bar{S}_{\rmE}}{\SE}, \;\;\;\;\;
 V \equiv t_IV^I, \;\;\;\;\;
 \Sgm \equiv t_I\Sgm^I.  \label{def:R_E}
\ee
The matrices~$t_I$ are the generators for the Abelian gauge groups. 

The above action is invariant under the gauge transformation: 
\bea
 H_{\rm odd} \toa e^{-\Lmd}H_{\rm odd}, \;\;\;\;\;
 H_{\rm even} \to e^\Lmd H_{\rm even}, \;\;\;\;\; (\Lmd \equiv t_I\Lmd^I) \nonumber\\
 V^I \toa V^I+\Lmd^I+\bar{\Lmd}^I, \;\;\;\;\;
 \Sgm^I \to \Sgm^I+\derE\Lmd^I, 
\eea
where $\Lmd^I$ are chiral superfields, and the other superfields are neutral. 
We should also note that (\ref{prev:S}) becomes the 5D SUGRA action 
in Refs.~\cite{Paccetti:2004ri,Abe:2004ar} 
with the norm function:~$\cN(X) = f_{IJ}X^IX^JX^T$ (the index~$T$ denotes the 5D vector multiplet 
originated from the 6D tensor multiplet) after the dimensional reduction. 

\ignore{
Note that we can absorb $U_E$ in (\ref{prev:S}) by redefining $V_E$ as
\be
 \tl{V}_{\rmE} \equiv \VE \UE, \label{V_E:absorb}
\ee
whose lowest component is 
\be
 \tl{V}_{\rmE}| = \frac{e^{(2)2}}{|E_4E_5|}. 
\ee
This field redefinition makes the transformation law considered in the next section simpler. 
Using (\ref{def:cV_T}), (\ref{prev:S}) is rewritten as
\bea
  \cL_{\rm H} \eql -\int\dr^4\tht\;2\tl{V}_{\rmE}^{1/2}
 \brkt{H_{\rm odd}^\dagger\tl{d}e^VH_{\rm odd}+H_{\rm even}^\dagger\tl{d}e^{-V}H_{\rm even}} 
 \nonumber\\
 &&+\sbk{\int\dr^2\tht\;\brc{H_{\rm odd}^t\tl{d}\brkt{\derE-\Sgm}H_{\rm even}
 -H_{\rm even}^t\tl{d}\brkt{\derE+\Sgm}H_{\rm odd}}+\hc}, \nonumber\\
 \cL_{\rm VT} \eql \int\dr^4\tht\;f_{IJ}\left[
 \brc{-2\Sgm^ID^\alp V^J\cY_{\rmT\alp}+\frac{1}{2}
 \brkt{\derE V^ID^\alp V^J-\derE D^\alp V^IV^J}\cY_{\rmT\alp}+\hc} \right.\nonumber\\
 &&\hspace{20mm}
 +\cV_{\rmT}\brkt{D^\alp V^I\cW_\alp^J
 +\frac{1}{2}V^ID^\alp\cW_\alp^J+\hc} \nonumber\\
 &&\hspace{20mm}
 +\frac{\cV_{\rmT}}{\tl{V}_{\rmE}}\left\{4\brkt{\bar{\der}_{\rmE}V^I-\bar{\Sgm}^I}\brkt{\derE V^J-\Sgm^J}
 -2\bar{\der}_{\rmE}V^I\derE V^J \right. \nonumber\\
 &&\hspace{30mm}\left.\left.
 +\frac{2\SE}{\bar{S}_{\rmE}}\Sgm^I\Sgm^J
 +\frac{2\bar{S}_{\rmE}}{\SE}\bar{\Sgm}^I\bar{\Sgm}^J\right\}\right]. 
 \label{prev:S:2}
\eea
}

We list the Weyl weights of the $\cN=1$ superfields in Table~\ref{Weyl_weight}. 
\begin{table}[t]
\begin{center}
\begin{tabular}{|c|c|c|c|c|c|c|c|c|c|c||c|c|c|c|} \hline
\multicolumn{5}{|c|}{$\bE$} & $\bH^A$ & \multicolumn{2}{|c|}{$\bV^I$} & 
\multicolumn{3}{|c||}{$\bT$} & \multicolumn{4}{|c|}{field strength} \\\hline
$U^\mu$ & $U^m$ & $\Psi_m^\alp$ & $\SE$ & $V_{\rm E}$ & 
$H^{\bar{A}}$ & $V^I$ & $\Sgm^I$ & $\Ups_{\rmT\alp}$ & $V_{\rmT m}$ & $\Sgm_{\rm T}$ 
& $\cW^I_\alp$ & $\cX_{\rmT}$ & $\cY_{\rmT\alp}$ & $\cV_{\rmT}$ \\\hline
0 & 0 & $-3/2$ & 0 & $-2$ & $3/2$ & 0 & 0 & $3/2$ & 0 & 0 & $3/2$ & 2 & $3/2$ & 0 \\\hline
\end{tabular}
\end{center}
\caption{The Weyl weights of the $\cN=1$ superfields. 
The 4D gravitational superfield~$U^\mu$ is explained in Appendix~\ref{4DSUGRAcouplings}, 
and the ``off-diagonal'' gravitational superfields~$U^m$ and $\Psi_m^\alp$ 
are introduced in Sec.~\ref{Xi-invariance} 
and Sec.~\ref{cov_derE}, respectively. }
\label{Weyl_weight}
\end{table}

\section{Diffeomorphism invariance in extra dimensions} \label{Xi-invariance}
Now we modify (\ref{prev:S}) by introducing the ``off-diagonal'' components of the 6D Weyl multiplet. 
For this purpose, we require the action to be
invariant under the diffeomorphism in the extra dimensions, \ie, $\dlt_\xi x^m=\xi^m$. 
The component field transformations are collected in Appendix~\ref{diffeo:comp}. 
It should be noted that we now have to discriminate the flat and the curved 4D indices 
even for the flat 4D background. 

\subsection{Hyper sector}
\subsubsection{Chiral superspace}
First, we focus on the chiral superspace in the hypersector. 

In the $\cN=1$ chiral superspace, the transformation parameters~$\xi^m$ are promoted 
to the chiral superfields as 
\be
 \Xi^m(x,\tht) = \xi^m(x)+ia^m(x)+\cO(\tht), 
\ee
where $a^m$ are real functions. 
From (\ref{diff:S_E}), (\ref{diff:H}) and (\ref{diff:Sgm}), the chiral superfields~$\SE$, 
$H_{\rm odd}$, $H_{\rm even}$ and $\Sgm^I$ transform as  
\bea
 \dlt_\Xi \SE \eql \Xi^m\der_m \SE
 +\frac{1}{2}\brkt{\der_4\Xi^4-\der_5\Xi^5+\frac{1}{\SE^2}\der_4\Xi^5-\SE^2\der_5\Xi^4}\SE, 
 \nonumber\\
 \dlt_\Xi H \eql \Xi^m\der_m H
 +\frac{1}{4}\brkt{\der_m\Xi^m+\frac{1}{\SE^2}\der_4\Xi^5+\SE^2\der_5\Xi^4}H, 
 \nonumber\\
 \dlt_\Xi\Sgm^I \eql \Xi^m\der_m\Sgm^I
 +\frac{1}{2}\brkt{\der_m\Xi^m-\frac{1}{\SE^2}\der_4\Xi^5-\SE^2\der_5\Xi^4}\Sgm^I, 
 \label{dlt_Xi:chiral}
\eea
where $H=H_{\rm odd},H_{\rm even}$. 
Because the first terms in the right-hand sides correspond to the shift of the coordinates~$x^m$, 
they have the universal structure for all the chiral superfields. 
In fact, noticing that 
\bea
 \dlt_\Xi(\derE H) \eql -(\dlt_\Xi \SE)\cO_{\rmE}H+\derE\brkt{\dlt_\Xi H} 
 \nonumber\\
 \eql \Xi^m\der_m\brkt{\derE H}
 +\frac{1}{4}\brkt{3\der_m\Xi^m-\frac{1}{\SE^2}\der_4\Xi^5-\SE^2\der_5\Xi^4}\derE H
 \nonumber\\
 &&+\frac{1}{4}\derE\brkt{\der_m\Xi^m+\frac{1}{\SE^2}\der_4\Xi^5+\SE^2\der_5\Xi^4}H, 
 \label{dlt_Xi:der_EH}
\eea
we can show that the chiral superspace part of the action~(\ref{prev:S}), \ie, 
the second line of $\cL_{\rm H}$, is invariant 
under (\ref{dlt_Xi:chiral}) up to total derivatives. 
\be
 \dlt_\Xi L_{\rm H}^{(1)} = \der_m\brkt{\Xi^m L_{\rm H}^{(1)}}, 
\ee
where
\be
 L_{\rm H}^{(1)} \equiv H_{\rm odd}^t\tl{d}\brkt{\derE-\Sgm}H_{\rm even}
 -H_{\rm even}^t\tl{d}\brkt{\derE+\Sgm}H_{\rm odd}. \label{def:L_H1}
\ee

\subsubsection{Full superspace}
Next we consider the invariance in the full superspace. 
There, terms originating from the shift of $x^m$ in the $\dlt_\Xi$-transformation should 
have a common form for all superfields. 
However, those for the chiral and the anti-chiral superfields have different forms. 
In order to accommodate them, we introduce the real superfields~$U^m$ ($m=4,5$), 
and introduce the operator~$\cP_U$ that shifts $x^m$ by $iU^m$. 
\bea
 \cP_U &:& x^m \to x^m+iU^m(x,\tht,\bar{\tht})
\eea
Then, for a chiral superfield~$\Phi$ (\ie, $\dlt_\Xi\Phi=\Xi^m\der_m\Phi+\cdots$), 
\be
 \hat{\Phi}(x,\tht,\bar{\tht}) \equiv \cP_U\Phi(x,\tht,\bar{\tht}) 
 = \Phi(x^\mu,x^m+iU^m(x,\tht,\bar{\tht}),\tht,\bar{\tht})  
\ee
transforms as~\footnote{
Note that $\der_m\hat{\Phi}=\widehat{\der_m\Phi}+i\der_m U^n\widehat{\der_n\Phi}$. 
}
\bea
 \dlt_\Xi\hat{\Phi} \eql \widehat{\dlt_\Xi\Phi}+i(\dlt_\Xi U^m)\widehat{\der_m\Phi} \nonumber\\
 \eql \hat{\Xi}^m\widehat{\der_m\Phi}+i(\dlt_\Xi U^m)\widehat{\der_m\Phi}+\cdots 
 \nonumber\\
 \eql  (\Re\hat{\Xi}^m)\der_m\hat{\Phi}+\cdots, \label{dlt_Xi:hatPhi}
\eea
if we assume that 
\be
 \dlt_\Xi U^m = -\Im\hat{\Xi}^m+(\Re\hat{\Xi}^n)\der_n U^m. \label{dlt_Xi:U^m}
\ee
Since $U^m$ transform nonlinearly, these correspond to the gauge fields 
for the $\dlt_\Xi$-transformation. 
The components of $U^m$ are identified as 
\bea
 U^m \eql (\tht\sgm^{\udl{\mu}}\bar{\tht})e_{\udl{\mu}}^{\;\;m}
 -\bar{\tht}^2(\tht\sgm^{\udl{\mu}}\bar{\psi}_{\udl{\mu}}^-)
 \brkt{e_{\udl{4}}^{\;\;m}+ie_{\udl{5}}^{\;\;m}} \nonumber\\
 &&+\tht^2\brkt{\bar{\tht}\bar{\sgm}^{\udl{\mu}}\psi_{\udl{\mu}}^-}
 \brkt{e_{\udl{4}}^{\;\;m}-ie_{\udl{5}}^{\;\;m}}+\cdots. 
 \label{comp:U^m}
\eea
Then, (\ref{dlt_Xi:U^m}) is consistent with the component transformation~(\ref{diff:e_bmu^m}).\footnote{
As we will explain in Sec.~\ref{comment:cP_U}, the $\tht\bar{\tht}$-component of $\Im\hat{\Xi}^m$ is 
$(\tht\sgm^{\udl{\mu}}\bar{\tht})\der_{\udl{\mu}}\xi^m$. 
}

For an anti-chiral superfield~$\bar{\Phi}$, 
\be
 \bar{\hat{\Phi}}(x,\tht,\bar{\tht}) \equiv \bar{\cP}_U\bar{\Phi}(x,\tht,\bar{\tht}) 
 = \bar{\Phi}(x^\mu,x^m-iU^m(x,\tht,\bar{\tht}),\tht,\bar{\tht}) 
\ee
transforms as
\be
 \dlt_\Xi\bar{\hat{\Phi}} = (\Re\hat{\Xi}^m)\der_m\bar{\hat{\Phi}}+\cdots, 
\ee
which has the same form as (\ref{dlt_Xi:hatPhi}). 

With the $\cP_U$ operation, (\ref{dlt_Xi:chiral}) becomes
\bea
 \dlt_\Xi\hat{S}_{\rmE} \eql (\Re\hat{\Xi}^m)\der_m\hat{S}_{\rmE}
 +\frac{1}{2}\brkt{\widehat{\der_4\Xi^4}-\widehat{\der_5\Xi^5}
 +\frac{1}{\hat{S}_{\rmE}^2}\widehat{\der_4\Xi^5}-\hat{S}_{\rmE}^2\widehat{\der_5\Xi^4}}\hat{S}_{\rmE}, 
 \nonumber\\
 \dlt_\Xi\hat{H} \eql (\Re\hat{\Xi}^m)\der_m\hat{H}
 +\frac{1}{4}\brkt{\widehat{\der_m\Xi^m}
 +\frac{1}{\hat{S}_{\rmE}^2}\widehat{\der_4\Xi^5}+\hat{S}_{\rmE}^2\widehat{\der_5\Xi^4}}\hat{H}, 
 \nonumber\\
 \dlt_\Xi\hat{\Sgm}^I \eql (\Re\hat{\Xi}^m)\der_m\hat{\Sgm}^I
 +\frac{1}{2}\brkt{\widehat{\der_m\Xi^m}
 -\frac{1}{\hat{S}_{\rmE}^2}\widehat{\der_4\Xi^5}
 -\hat{S}_{\rmE}^2\widehat{\der_5\Xi^4}}\hat{\Sgm}^I. \label{dlt_Xi:hatchiral}
\eea

From (\ref{sf:components}) and (\ref{diff:V}), 
the $\dlt_\Xi$-transformation of the vector superfield~$V^I$ is found to be 
\be
 \dlt_\Xi V^I = (\Re\hat{\Xi}^m)\der_m V^I. \label{dlt_Xi:V^I}
\ee

Therefore, the combination
\be
 L_{\rm H}^{(2)} \equiv \hat{H}_{\rm odd}^\dagger\tl{d}e^V\hat{H}_{\rm odd}
 +\hat{H}_{\rm even}^\dagger\tl{d}e^{-V}\hat{H}_{\rm even} \label{def:L_H^2}
\ee 
in the first line of $\cL_{\rm H}$ transforms as
\be
 \dlt_\xi L_{\rm H}^{(2)} = (\Re\hat{\Xi}^m)\der_m L_{\rm H}^{(2)}
 +\frac{1}{2}\Re\brkt{\widehat{\der_m\Xi^m}+\frac{1}{\hat{S}_{\rm E}^2}\widehat{\der_4\Xi^5}
 +\hat{S}_{\rm E}^2\widehat{\der_5\Xi^4}}L_{\rm H}^{(2)}. 
\ee
As for the factor in front of $L_{\rm H}^{(2)}$ in $\cL_{\rm H}$,  
we should note that the combination~$V_{\rmE}\UE$ transforms as 
\be
 \dlt_\Xi\brkt{V_{\rmE}\UE} = (\Re\hat{\Xi}^m)\der_m\brkt{V_{\rmE}\UE}
 +\Re\brkt{\widehat{\der_m\Xi^m}
 -\frac{1}{\hat{S}_{\rmE}^2}\widehat{\der_4\Xi^5}-\hat{S}_{\rmE}^2\widehat{\der_5\Xi^4}}
 \brkt{V_{\rmE}\UE}, 
 \label{dlt_Xi:tlV_E}
\ee
which is consistent with (\ref{diff:tlV_E}). 
This transformation law is derived from (\ref{dlt_Xi:V_E}) and (\ref{dlt_Xi:U_E}) 
explained later. 

Consider the Jacobian for $\cP_U$, which is calculated as
\bea
 J_\cP \defa {\rm sdet}\,\brkt{\frac{\der(x^\mu,x^m+iU^m(x,\tht),\tht^\alp,\bar{\tht}_{\dalp})}
 {\der(x^\nu,x^n,\tht^\bt,\bar{\tht}_{\dbt})}} \nonumber\\
 \eql 1+i\der_m U^m-\der_4U^4\der_5U^5+\der_4U^5\der_5U^4,  \label{def:J_cP}
\eea
which satisfies 
\bea
 \int\dr^6xd^4\tht\;J_\cP\hat{\Phi} \eql \int\dr^6xd^4\tht\;\Phi = 0, 
 \label{prop:linear}
%
\eea
for a chiral superfield~$\Phi$. 
After some calculations, we can show that $J_\cP$ transforms as
\be
 \dlt_\Xi J_\cP = \der_m\brc{(\Re\hat{\Xi}^m)J_\cP}-\widehat{\der_m\Xi^m}J_\cP. 
 \label{dlt_Xi:J_cP}
\ee
Then, we obtain
\bea
 \dlt_\Xi\abs{J_\cP} \eql \Re\hat{\Xi}^m\der_m\abs{J_\cP}
 +\Re\brkt{\der_m\hat{\Xi}^m-\widehat{\der_m\Xi^m}}\abs{J_\cP}. 
\eea
Combining these transformation laws, we find 
\bea
 \dlt_\Xi\brkt{\abs{J_\cP}V_{\rm E}^{1/2}\UE^{1/2} L_{\rm H}^{(2)}} 
 \eql \der_m\brkt{\Re\hat{\Xi}^m\abs{J_\cP}V_{\rm E}^{1/2}\UE^{1/2} L_{\rm H}^{(2)}}. 
\eea

\subsubsection{Comment on $\bdm{\cP_U}$} \label{comment:cP_U}
Here, we give a comment on the operator~$\cP_U$. 
Let us consider a chiral superfield~$\Phi$ whose components are given by 
\be
 \Phi = \phi+\tht\psi+\tht^2 F
 +i(\tht\sgm^{\udl{\mu}}\bar{\tht})\der_\mu\phi 
 -\frac{i}{2}\tht^2\der_\mu\psi\sgm^{\udl{\mu}}\bar{\tht}
 +\frac{1}{4}\tht^2\bar{\tht}^2\Box_4\phi, 
\ee
where $\Box_4\equiv\der_\mu\der^\mu$. 
After the $\cP_U$ operation, this becomes
\bea
 \hat{\Phi}(x,\tht) \eql \Phi(x,\tht)+iU^m\der_m\Phi+\cO(U^2) \nonumber\\
 \eql \phi+\tht\psi+\tht^2F+i(\tht\sgm^{\udl{\mu}}\bar{\tht})
 \brkt{\der_\mu\phi+e_{\udl{\mu}}^{\;\;m}\der_m\phi} \nonumber\\
 &&-\frac{i}{2}\tht^2\brkt{\der_\mu\psi+e_{\udl{\mu}}^{\;\;m}\der_m\psi}\sgm^{\udl{\mu}}\bar{\tht}
 +\cdots. 
\eea
Namely, the operator~$\cP_U$ replaces the derivative~$\der_\mu$ appearing 
in the components with 
\bea
 \der_{\udl{\mu}} \eql e_{\udl{\mu}}^{\;\;N}\der_N 
 = e_{\udl{\mu}}^{\;\;\nu}\der_\nu+e_{\udl{\mu}}^{\;\;m}\der_m \nonumber\\
 \eql \der_\mu+e_{\udl{\mu}}^{\;\;m}\der_m. 
\eea
We have dropped the fluctuation of $e_\mu^{\;\;\udl{\nu}}$ around the background~$\dlt_\mu^{\;\;\udl{\nu}}$, 
and terms beyond linear in the ``off-diagonal'' components of the sechsbein. 
Recall that the index of $\sgm^{\udl{\mu}}$ is the flat one. 
So the 4D indices contracted with it should also be the flat ones. 
In higher-dimensional SUGRA, this means that terms 
involving the ``off-diagonal'' components of the vielbein must be incorporated, 
which are missing in the original superfield~$\Phi$. 
The operator~$\cP_U$ provides such missing terms. 

For later convenience, we ``covariantize'' the spinor derivatives~$D_\alp$ and $\bar{D}_{\dalp}$ as 
\be
 D_\alp^\cP \equiv \bar{\cP}_U D_\alp\bar{\cP}_U^{-1}, \;\;\;\;\;
 \bar{D}_{\dalp}^\cP \equiv \cP_U\bar{D}_{\dalp}\cP_U^{-1}. \label{def:hatD}
\ee
Then, we can also see the same effect of $\cP_U$ in the $\cN=1$ SUSY algebra. 
\bea
 \brc{D_\alp^\cP,\bar{D}_{\dalp}^\cP} \eql \left\{
 D_\alp+iD_\alp U^m\der_m+\cO(U^2), 
 \bar{D}_{\dalp}-i\bar{D}_{\dalp}U^n\der_n+\cO(U^2)\right\} \nonumber\\
 \eql \brc{D_\alp,\bar{D}_{\dalp}}-i\sbk{D_\alp,\bar{D}_{\dalp}}U^m\der_m+\cO(U^2) \nonumber\\
 \eql -2i\sgm_{\alp\dalp}^{\udl{\mu}}\der_\mu
 -i\brkt{2\sgm_{\alp\dalp}^{\udl{\mu}}e_{\udl{\mu}}^{\;\;m}\der_m+\cdots}+\cO(U^2) \nonumber\\
 \eql -2i\sgm^{\udl{\mu}}_{\alp\dalp}\der_{\udl{\mu}}+\cdots, 
 \label{DbD:hat}
\eea
where $\cO(U^2)$ denotes terms beyond linear in $U^m$.

\subsection{Vector-tensor sector}
\subsubsection{Field strength superfields}
From (\ref{dlt_Xi:V^I}), we can show that~\footnote{
Notice that $\cP_U^{-1}$ is different from $\bar{\cP}_U$ because 
\bea
 \cP_U^{-1}x^m \eql x^m-iU^m(\cP_U^{-1}x,\tht) \nonumber\\
 \eql x^m-iU^m(x,\tht)-U^n(x,\tht)\der_n U^m(x,\tht)+\cdots. 
\eea
}
\bea
 \dlt_\Xi\brkt{\cP_U^{-1}V^I} \eql \Xi^m\der_m\brkt{\cP_U^{-1}V^I}, \nonumber\\
 \dlt_\Xi\brkt{\bar{\cP}_U^{-1}V^I} \eql \bar{\Xi}^m\der_m\brkt{\bar{\cP}_U^{-1}V^I}. 
 \label{dlt_Xi:invPV}
\eea
%
Hence, if we modify the field strength superfield~$\cW_\alp^I$ 
in (\ref{def:cW_alp^I:1}) as
\bea
 \hat{\cW}_\alp^I \equiv -\frac{1}{4}(\bar{D}^\cP)^2 D_\alp^\cP V^I, 
 \label{def:cW:naive}
\eea
it transforms as
\be
 \dlt_\Xi\hat{\cW}_\alp^I = (\Re\hat{\Xi}^m)\der_m\hat{\cW}_\alp^I, \label{dlt_Xi:cW}
\ee
which is consistent with the component transformation. 
However, this is not gauge-invariant under
\be
 \dlt_\Lmd V^I = \hat{\Lmd}^I+\bar{\hat{\Lmd}}^I 
\ee
because 
\bea
 \dlt_\Lmd\hat{\cW}_\alp^I \eql -\frac{1}{4}\brkt{\bar{D}^\cP}^2D_\alp^\cP\hat{\Lmd}^I 
 \nonumber\\ 
 \eql \cP_U\brkt{-\frac{i}{2}\bar{D}^2D_\alp U^m\der_m\Lmd}+\cO(U^2). 
 \label{dlt_Lmd:cW:naive}
\eea
This stems from the fact that $\cW_\alp^I$ should include the field strength~$F_{\udl{\mu}\udl{\nu}}$, and 
\bea
 F_{\udl{\mu}\udl{\nu}} \eql e_{\udl{\mu}}^{\;\;L}e_{\udl{\nu}}^{\;\;P}\der_L A_P-(\mu\exch\nu) 
 \nonumber\\
 \eql e_{\udl{\mu}}^{\;\;L}\brkt{\der_L A_{\udl{\nu}}
 -\der_L e_{\udl{\nu}}^{\;\;P}A_P}-\brkt{\mu\exch\nu} \nonumber\\ 
 \eql \brkt{\der_{\udl{\mu}}A_{\udl{\nu}}-\der_{\udl{\nu}}A_{\udl{\mu}}}
 -\brkt{\der_{\udl{\mu}}e_{\udl{\nu}}^{\;\;n}-\der_{\udl{\nu}}e_{\udl{\mu}}^{\;\;n}}A_n+\cdots, 
 \label{expand:Fcomp}
\eea
where the ellipsis denotes terms beyond the linear order 
in the ``off-diagonal'' components~$\{e_m^{\;\;\udl{\nu}},e_{\udl{\mu}}^{\;\;n}\}$,
or terms involving the fluctuation of $e_\mu^{\;\;\udl{\nu}}$. 
The superfield defined in (\ref{def:cW:naive}) only contains the first term in (\ref{expand:Fcomp}). 
Thus, we have to modify (\ref{def:cW:naive}) by adding terms that depend on 
$U^m$ and $\Sgm^I$, in order to cancel the variation~(\ref{dlt_Lmd:cW:naive}). 
The identification of the additional terms is left 
for the subsequent paper, in which the gauge group is extended to non-Abelian, 
but such correction terms should be determined so that  
the transformation law~(\ref{dlt_Xi:cW}) is maintained. 

\ignore{
In order to include the second term, we should modify (\ref{def:cW:naive}) as
\bea
 \hat{\cW}_\alp^I \defa -\frac{1}{4}\brkt{\bar{D}^\cP}^2D_\alp^\cP\left[V^I
 -\frac{U^4}{\UE}\brc{\bar{S}_{\rm E}\Sgm^I-\SE\brkt{\bar{\der}_{\rm E}V^I-\bar{\Sgm}^I}}
 \right.\nonumber\\
 &&\hspace{30mm}\left. 
 -\frac{U^5}{\UE}\brc{\frac{\Sgm^I}{\bar{S}_{\rm E}}
 -\frac{1}{\SE}\brkt{\bar{\der}_{\rm E}V^I-\bar{\Sgm}^I}}
 \right]+\cO(U^2). \label{def:cW}
\eea
\ignore{
Here, note that 
\bea
 &&-\frac{1}{4}\brkt{\bar{D}^\cP}^2D^\cP_\alp U^n 
 = i\brkt{\sgm^{\udl{\mu}\udl{\nu}}\tht}_\alp\brkt{\der_{\udl{\mu}}e_{\udl{\nu}}^{\;\;n}
 -\der_{\udl{\nu}}e_{\udl{\mu}}^{\;\;n}}+\cdots, 
 \nonumber\\
 &&\left.\frac{1}{\UE}\brc{\bar{S}_E\Sgm^I-\brkt{\bar{\der}_EV^I-\bar{\Sgm}^I}}\right| 
 = -A_4, \nonumber\\
 &&\left.\frac{1}{U_E^2}\brc{\frac{\Sgm^I}{\bar{S}_E}-\frac{1}{S_E}\brkt{\bar{\der}_EV^I-\bar{\Sgm}^I}}
 \right| = -A_5. 
\eea
}
In fact, since 
\bea
 \dlt_\Lmd\sbk{\frac{1}{\UE}\brc{\bar{S}_{\rm E}\Sgm^I
 -\SE\brkt{\bar{\der}_{\rm E}V^I-\bar{\Sgm}^I}}} 
 \eql 2i\der_4\Lmd+\cO(U^m), \nonumber\\
 \dlt_\Lmd\sbk{\frac{1}{\UE}\brc{\frac{\Sgm^I}{\bar{S}_{\rm E}}
 -\frac{1}{\SE}\brkt{\bar{\der}_{\rm E}V^I-\bar{\Sgm}^I}}} \eql 2i\der_5\Lmd+\cO(U^m), 
\eea
we can see that
\be
 \dlt_\Lmd\hat{\cW}_\alp^I = \cO(U^2). 
\ee
The higher order terms~$\cO(U^2)$ in (\ref{def:cW}) should be determined 
so that the transformation law~(\ref{dlt_Xi:cW}) is maintained. 
Thus, the definition in (\ref{def:cW:naive}) is not the appropriate one. 
In order to find the correct definition of $\hat{\cW}_\alp^I$, we first note that 
(\ref{def:cW_alp^I:1}) can be rewritten as
\be
 t_I\cW_\alp^I = \frac{1}{4}\bar{D}^2\brkt{e^V D_\alp e^{-V}}, 
\ee
which is also applicable in the nonabelian gauge theory. 
Now we define $\hat{\cW}_\alp^I$ by replacing the spinor derivatives 
with the ``covariant'' one~$D_\alp^\cP$ and $\bar{D}_{\dalp}^\cP$ in this expression. 
\bea
 t_I\hat{\cW}_\alp^I \defa \frac{1}{4}(\bar{D}^\cP)^2
 \brkt{e^V D_\alp^\cP e^{-V}} \nonumber\\
 \eql \frac{1}{4}\cP_U\bar{D}^2\brkt{\cP^{-1}_U e^V\bar{\cP}_UD_\alp\bar{\cP}_U^{-1}e^{-V}}
\eea
}

Next we consider the tensor multiplet. 
The $\dlt_\Xi$-transformations of $\Ups_{\rmT\alp}$, $V_{\rmT m}$ and $\Sgm_{\rmT}$ are found 
from (\ref{diffeo}) and (\ref{dlt_xi:Bs}) as
\bea
 \dlt_\Xi\Ups_{\rmT\alp} \eql \Xi^m\der_m\Ups_{\rmT\alp}, \nonumber\\
 \dlt_\Xi V_{\rmT m} \eql \Re\hat{\Xi}^n\der_n V_{\rmT m}
 +(\Re\der_m\hat{\Xi}^n) V_{\rmT n},   \nonumber\\
 \dlt_\Xi\Sgm_{\rmT} \eql \der_m\brkt{\Xi^m\Sgm_{\rmT}}. 
 \label{dlt_Xi:V_Tm}
\eea
The definition of the field strength~$\cX_{\rmT}$ in (\ref{def:cXcY}) is modified as
\be
 \cX_{\rmT} \equiv \frac{1}{2}\Im\brkt{D^{\cP\alp}\Ups_{\rmT\alp}}. 
\ee
Then, it transforms as
\be
 \dlt_\Xi\cX_{\rmT} = \Re\hat{\Xi}^m\der_m\cX_{\rmT}.  \label{dlt_Xi:cX_T}
\ee

The second term in $\dlt_\Xi V_{\rmT m}$ exists because 
$V_{\rmT m}$ has an external index~$m$. 
Thus we extend the operator~$\cP_U$ as follows. 
For a chiral superfield~$\Phi_m$, we define the operator~$\cQ_U$ as~\footnote{
The operators~$\cP_U$ and $\cQ_U$ are understood as $e^{i\cL_U}$, 
where $\cL_U$ is the Lie derivative along $U^m$. 
} 
\be
 \cQ_U\Phi_m = \hat{\Phi}_m+i\der_m U^n\hat{\Phi}_n. 
\ee
Since $\Phi_m$ has an external index~$m$, its $\dlt_\Xi$-transformation has a form of 
\be
 \dlt_\Xi\Phi_m = \Xi^n\der_n\Phi_m+\der_m\Xi^n\Phi_n+\cdots, 
\ee
Then we can show that
\be
 \dlt_\Xi(\cQ_U\Phi_m) = \Re\hat{\Xi}^n\der_n(\cQ_U\Phi_m)
 +(\Re\der_m\hat{\Xi}^n)\cQ_U\Phi_n+\cdots. 
\ee
Note that this has the same form as $\dlt_\Xi V_{\rmT m}$ in (\ref{dlt_Xi:V_Tm}). 
Hence, it follows that~\footnote{
Specifically, $\cQ_U^{-1}V_{\rmT m}$ is 
\bea
 \cQ_U^{-1}V_{\rmT m}(x) \eql V_{\rmT m}(\cP_U^{-1}x)
 -i(\cP_U^{-1}\der_m U^n)\brc{\cQ_U^{-1}V_{\rmT n}}(x) \nonumber\\
 \eql V_{\rmT m}(\cP_U^{-1}x)-i(\cP_U^{-1}\der_m U^n)V_{\rmT n}(\cP_U^{-1}x) \nonumber\\
 &&-(\cP_U^{-1}\der_m U^n)(\cP_U^{-1}\der_n U^l)V_{\rmT l}(\cP_U^{-1}x)+\cdots. 
\eea
}
\bea
 \dlt_\Xi(\cQ_U^{-1}V_{\rmT m}) \eql \Xi^n\der_n(\cQ_U^{-1}V_{\rmT m})
 +\der_m\Xi^n(\cQ_U^{-1}V_{\rmT n}), \nonumber\\
 \dlt_\Xi(\bar{\cQ}_U^{-1}V_{\rmT m}) \eql \bar{\Xi}^n\der_n(\bar{\cQ}_U^{-1}V_{\rmT m})
 +\der_m\bar{\Xi}^n(\bar{\cQ}_U^{-1}V_{\rmT n}). 
\eea
Making use of these properties, $\cW_{\rmT m\alp}$ in (\ref{def:cW_Tmalp}) should be modified as
\bea
 \cW_{\rmT m\alp} \eql -\frac{1}{4}\bar{D}^2\cQ_U^{-1}\bar{\cQ}_U D_\alp\bar{\cQ}_U^{-1}V_{\rmT m} 
 \nonumber\\
 \eql \cQ_U^{-1}\brc{-\frac{1}{4}(\bar{D}^{\cQ})^2D^{\cQ}_\alp V_{\rmT m}}, 
\eea
where
\be
 D_\alp^\cQ \equiv \bar{\cQ}_U D_\alp\bar{\cQ}_U^{-1}, \;\;\;\;\;
 \bar{D}_{\dalp}^\cQ \equiv \cQ_U\bar{D}_{\dalp}\cQ_U^{-1}. 
\ee
Then, it transforms as
\be
 \dlt_\Xi\cW_{\rmT m\alp} = \Xi^n\der_n\cW_{\rmT m\alp}+\der_m\Xi^n\cW_{\rmT n\alp}, 
\ee
which leads to
\bea
 \dlt_\Xi\brkt{\frac{\cW_{\rmT 4\alp}}{\SE}+\SE\cW_{\rmT 5\alp}} 
 \eql \Xi^m\der_m\brkt{\frac{\cW_{\rmT 4\alp}}{\SE}+\SE\cW_{\rmT 5\alp}} \nonumber\\
 &&+\frac{1}{2}\brkt{\der_m\Xi^m+\frac{1}{\SE^2}\der_4\Xi^5+\SE^2\der_5\Xi^4}
 \brkt{\frac{\cW_{\rmT 4\alp}}{\SE}+\SE\cW_{\rmT 5\alp}} \nonumber\\
 &&-\brkt{\frac{1}{\SE^2}\der_4\Xi^5-\SE^2\der_5\Xi^4}
 \brkt{\frac{\cW_{\rmT 4\alp}}{\SE}-\SE\cW_{\rmT 5\alp}}. \label{dlt_Xi:cY1}
\eea
From (\ref{dlt_Xi:V_Tm}), we can show that
\bea
 \dlt_\Xi\brkt{\SE\cO_{\rmE}\Ups_{\rmT\alp}} 
 \eql \SE\cO_{\rmE}\brkt{\dlt_\Xi\Ups_{\rmT\alp}}
 -\frac{\dlt_\Xi\SE}{\SE}\derE\Ups_{\rmT\alp} \nonumber\\
 \eql \Xi^m\der_m\brkt{\SE\cO_{\rmE}\Ups_{\rmT\alp}}
 +\frac{1}{2}\brkt{\der_m\Xi^m+\frac{1}{\SE^2}\der_4\Xi^5+\SE^2\der_5\Xi^4}
 \SE\cO_{\rmE}\Ups_{\rmT\alp} \nonumber\\
 &&-\brkt{\frac{1}{\SE^2}\der_4\Xi^5-\SE^2\der_5\Xi^4}\derE\Ups_{\rmT\alp}. 
 \label{dlt_Xi:cY2}
\eea
Summing (\ref{dlt_Xi:cY1}) and (\ref{dlt_Xi:cY2}), we obtain the $\dlt_\Xi$-transformation 
of $\cY_{\rmT\alp}$ defined in (\ref{def:cXcY}) as
\be
 \dlt_\Xi\cY_{\rmT\alp} = \Xi^m\der_m\cY_{\rmT\alp}
 +\frac{1}{2}\brkt{\der_m\Xi^m+\frac{1}{\SE^2}\der_4\Xi^5+\SE^2\der_5\Xi^4}\cY_{\rmT\alp}. 
 \label{dlt_Xi:cY}
\ee
We have used the constraint~(\ref{Tconstraints}). 

\ignore{
The other field strength~$\cX_{\rmT}$ in (\ref{def:cXcY}) should be modified as
\be
 \cX_{\rmT} \equiv \frac{1}{2}\Im\brkt{\hat{D}^\alp\hat{\Ups}_{\rmT\alp}}. 
\ee
Then, this transforms as
\be
 \dlt_\Xi\cX_{\rmT} = \Re\hat{\Xi}^m\der_m\cX_{\rmT}, 
\ee
which is consistent with the fifth transformation in (\ref{diffeo}). 
}

From (\ref{dlt_Xi:V_Tm}), we also obtain 
\bea
 \dlt_\Xi\hat{\Sgm}_{\rmT} \eql \Re\hat{\Xi}^m\der_m\hat{\Sgm}_{\rmT}
 +\widehat{\der_m\Xi^m}\hat{\Sgm}_{\rmT},  \nonumber\\
 \dlt_\Xi\brkt{\der_4 V_{\rmT 5}-\der_5 V_{\rmT 4}} 
 \eql \der_4\brkt{\dlt_\Xi V_{\rmT 5}}-\der_5\brkt{\dlt_\Xi V_{\rmT 4}} \nonumber\\
 \eql \der_m\brc{\Re\hat{\Xi}^m\brkt{\der_4 V_{\rmT 5}-\der_5 V_{\rmT 4}}}, \nonumber\\
 \dlt_\Xi\brkt{J_\cP\hat{\Sgm}_{\rmT}} 
 \eql \brkt{\dlt_\Xi J_\cP}\hat{\Sgm}_{\rmT}+J_\cP\dlt_\Xi\hat{\Sgm}_{\rmT} \nonumber\\
 \eql \der_m\brc{(\Re\hat{\Xi}^m) J_\cP\hat{\Sgm}_{\rmT}}.  
\eea
Therefore, if we modify the definition of $\cV_{\rmT}$ in (\ref{def:cV_T}) as
\be
 \cV_{\rmT} \equiv \der_4 V_{\rmT 5}-\der_5 V_{\rmT 4}+J_\cP\hat{\Sgm}_{\rmT}
 +\bar{J}_\cP\bar{\hat{\Sgm}}_{\rmT}, 
\ee
we find that
\be
 \dlt_\Xi\brkt{\cV_{\rmT}} = \der_m\brkt{\Re\hat{\Xi}^m\cV_{\rmT}}. 
 \label{dlt_Xi:cV_T}
\ee

Recall that $\VE=\cV_{\rmT}/\cX_{\rmT}$ from (\ref{def:cV_T}). 
Thus, from (\ref{dlt_Xi:cX_T}) and (\ref{dlt_Xi:cV_T}), we obtain
\bea
 \dlt_\Xi\VE \eql \dlt_\Xi\brkt{\frac{\cV_{\rmT}}{\cX_{\rmT}}}
 = \der_m\brkt{\Re\hat{\Xi}^m\frac{\cV_{\rmT}}{\cX_{\rmT}}} \nonumber\\
 \eql \Re\hat{\Xi}^m\der_m\brkt{\frac{\cV_{\rmT}}{\cX_{\rmT}}}
 +\brkt{\Re\widehat{\der_m\Xi^m}-\der_m U^n\Im\widehat{\der_n\Xi^m}}
 \frac{\cV_{\rmT}}{\cX_{\rmT}}, \label{dlt_Xi:V_E}
\eea
which is consistent with (\ref{dlt_xi:e^2}). 
However, this and (\ref{dlt_Xi:hatchiral}) are not consistent with (\ref{dlt_Xi:tlV_E}). 
Hence, we modify the definition of $\UE$ given in (\ref{def:R_E}) 
in such a way that $\VE \UE$ transforms as (\ref{dlt_Xi:tlV_E}). 
We modify $\UE$ as
\be
 \UE \equiv \frac{1}{2}\Im\brkt{J_S^{(2)}\frac{\bar{\hat{S}}_{\rm E}}{\hat{S}_{\rm E}}
 -J_S^{(1)}\frac{\hat{S}_{\rm E}}{\bar{\hat{S}}_{\rm E}}}, \label{def:U_E}
\ee
where
\bea
 J_S^{(1)} \defa 1+i\brkt{\der_4U^4-\der_5U^5}
 -2i\bar{\hat{S}}_{\rm E}^2\der_5 U^4+\cO(U^2), \nonumber\\
 J_S^{(2)} \defa 1-i\brkt{\der_4 U^4-\der_5U^5}
 -\frac{2i}{\bar{\hat{S}}_{\rm E}^2}\der_4 U^5+\cO(U^2). \label{def:J_S}
\eea
The higher order terms~$\cO(U^2)$ are determined so that $J_S^{(1)}$ and $J_S^{(2)}$ transform as
\bea
 \dlt_\Xi J_S^{(1)} \eql \Re\hat{\Xi}^m\der_m J_S^{(1)}
 -i\brc{\Im\brkt{\widehat{\der_4\Xi^4}-\widehat{\der_5\Xi^5}}
 -2\bar{\hat{S}}_{\rm E}^2\Im\widehat{\der_5\Xi^4}}J_S^{(1)} \nonumber\\
 &&-\brc{\der_m U^n\Im\widehat{\der_n\Xi^m}
 -2i|\hat{S}_{\rm E}|^2\brkt{\frac{\UE}{J_S^{(1)}}
 -\Im\frac{\bar{\hat{S}}_{\rm E}}{\hat{S}_{\rm E}}}\widehat{\der_5\Xi^4}}J_S^{(1)}, \nonumber\\
 \dlt_\Xi J_S^{(2)} \eql \Re\hat{\Xi}^m\der_m J_S^{(2)}
 +i\brc{\Im\brkt{\widehat{\der_4\Xi^4}-\widehat{\der_5\Xi^5}}
 +\frac{2}{\bar{\hat{S}}_{\rm E}^2}\Im\widehat{\der_4\Xi^5}}J_S^{(2)} \nonumber\\
 &&-\brc{\der_m U^n\Im\widehat{\der_n\Xi^m}
 +\frac{2i}{|\hat{S}_{\rm E}|^2}\brkt{\frac{\UE}{J_S^{(2)}}
 -\Im\frac{\bar{\hat{S}}_{\rm E}}{\hat{S}_{\rm E}}}\widehat{\der_4\Xi^5}}J_S^{(2)}. 
\eea
These lead to 
\bea
 \dlt_\Xi\brkt{J_S^{(1)}\frac{\hat{S}_{\rm E}}{\bar{\hat{S}}_{\rm E}}} 
 \eql \Re\hat{\Xi}^m\der_m\brkt{J_S^{(1)}\frac{\hat{S}_{\rm E}}{\bar{\hat{S}}_{\rm E}}} 
 +2i\hat{S}_{\rm E}^2\widehat{\der_5\Xi^4}\UE 
 \nonumber\\
 &&+\brc{-\Re\brkt{i\der_m U^n\widehat{\der_n\Xi^m}}
 +i\Im\brkt{\frac{1}{\hat{S}_{\rm E}^2}\widehat{\der_4\Xi^5}
 +\hat{S}_{\rm E}^2\widehat{\der_5\Xi^4}}}J_S^{(1)}\frac{\hat{S}_{\rm E}}{\bar{\hat{S}}_{\rm E}}, \nonumber\\
 \dlt_\Xi\brkt{J_S^{(2)}\frac{\bar{\hat{S}}_{\rm E}}{\hat{S}_{\rm E}}} 
 \eql \Re\hat{\Xi}^m\der_m\brkt{J_S^{(2)}\frac{\bar{\hat{S}}_{\rm E}}{\hat{S}_{\rm E}}} 
 -\frac{2i}{\hat{S}_{\rm E}^2}\widehat{\der_4\Xi^5}\UE \nonumber\\
 &&+\brc{-\Re\brkt{i\der_m U^n\widehat{\der_n\Xi^m}}
 +i\Im\brkt{\frac{1}{\hat{S}_{\rm E}^2}\widehat{\der_4\Xi^5}
 +\hat{S}_{\rm E}^2\widehat{\der_5\Xi^4}}}J_S^{(2)}\frac{\bar{\hat{S}}_{\rm E}}{\hat{S}_{\rm E}}. 
 \label{dlt_Xi:J_SSbS}
\eea
As a result, $\UE$ transforms as
\be
 \dlt_\Xi \UE = \Re\hat{\Xi}^m\der_m \UE
 -\Re\brkt{\frac{1}{\hat{S}_{\rm E}^2}\widehat{\der_4\Xi^5}
 +\hat{S}_{\rm E}^2\widehat{\der_5\Xi^4}
 +i\der_m U^n\widehat{\der_n\Xi^m}}\UE. \label{dlt_Xi:U_E}
\ee
From (\ref{dlt_Xi:V_E}) and (\ref{dlt_Xi:U_E}), we certainly obtain the transformation law~(\ref{dlt_Xi:tlV_E}).

\subsubsection{Invariance of action}
Let us first consider the $\dlt_\Xi$-invariance of the first line of $\cL_{\rm VT}$ in (\ref{prev:S}). 
If we define
\be
 \der_{\rmE}^\cP \equiv \cP_U\der_{\rmE}\cP_U^{-1}, 
\ee
we find that 
\be
 \dlt_\Xi\brkt{\der_{\rmE}^\cP V^I} = (\Re\hat{\Xi}^m)\der_m\brkt{\der_{\rmE}^\cP V^I} 
 +\frac{1}{2}\brkt{\widehat{\der_m\Xi^m}
 -\frac{1}{\hat{S}_{\rmE}^2}\widehat{\der_4\Xi^5}
 -\hat{S}_{\rmE}^2\widehat{\der_5\Xi^4}}\der_{\rmE}^\cP V^I. 
 \label{dlt_Xi:derEcPV}
\ee
This is the same transformation law as that of $\hat{\Sgm}^I$. 
Similarly, $\der_{\rmE}^{\cP}D^\cP_\alp V^I$ also has 
the same transformation law.  
Combining these properties with (\ref{dlt_Xi:cY}), we can show that  
\be
 \dlt_\Xi \brkt{L_{\rm V}^{(1)\alp}\hat{\cY}_{\rm T\alp}} 
 = (\Re\hat{\Xi}^m)\der_m \brkt{L_{\rm V}^{(1)\alp}\hat{\cY}_{\rm T\alp}}
 +\widehat{\der_m\Xi^m}\brkt{L_{\rm V}^{(1)\alp}\hat{\cY}_{\rm T\alp}}, 
\ee
where
\be
 L_{\rm V}^{(1)\alp} \equiv f_{IJ}\brc{-2\hat{\Sgm}^I D^{\cP\alp} V^J
 +\frac{1}{2}\brkt{\der_{\rmE}^\cP V^I D^{\cP\alp} V^J
 -\der_{\rmE}^\cP D^{\cP\alp} V^IV^J}}. \label{def:L_V^1}
\ee

Recalling (\ref{dlt_Xi:J_cP}), we find that
\be
 \dlt_\Xi\brkt{J_\cP L_{\rm V}^{(1)\alp}\hat{\cY}_{\rm T\alp}} 
 = \der_m\brc{(\Re\hat{\Xi}^m) J_\cP L_{\rm V}^{(1)\alp}\hat{\cY}_{\rm T\alp}}. 
\ee

Next, consider the second line of $\cL_{\rm VT}$. 
Since the combination 
\be
 L_{\rm V}^{(2)} \equiv f_{IJ}\brkt{D^{\cP\alp} V^I\hat{\cW}_\alp^J
 +\frac{1}{2}V^I D^{\cP\alp}\hat{\cW}_\alp^J+\hc} 
\ee
transforms as
\be
 \dlt_\Xi L_{\rm V}^{(2)} = (\Re\hat{\Xi}^m)\der_m L_{\rm V}^{(2)}, 
\ee
we find that
\bea
 \dlt_\Xi\brkt{\cV_{\rmT} L_{\rm V}^{(2)}} 
 \eql \der_m\brkt{\Re\hat{\Xi}^m\cV_{\rmT}}L_{\rm V}^{(2)}
 +\cV_{\rmT}\cdot(\Re\hat{\Xi}^m)\der_m L_{\rm V}^{(2)} \nonumber\\
 \eql \der_m\brkt{\Re\hat{\Xi}^m\cV_{\rmT} L_{\rm V}^{(2)}}. 
\eea

As for the third line of $\cL_{\rm VT}$, the combination 
\be
 L_{\rm V}^{(3)} \equiv f_{IJ}\brc{4\brkt{\der_{\rmE}^\cP V^I-\hat{\Sgm}^I}^\dagger
 \brkt{\der_{\rmE}^\cP V^J-\hat{\Sgm}^J}
 -2\brkt{\der_{\rmE}^\cP V^I}^\dagger \der_{\rmE}^\cP V^J} \label{def:L_V^3}
\ee
transforms as 
\bea
 \dlt_\Xi L_{\rm V}^{(3)} \eql (\Re\hat{\Xi}^m)\der_m L_{\rm V}^{(3)}
 +\Re\brkt{\widehat{\der_m\Xi^m}-\frac{1}{\hat{S}_{\rmE}^2}\widehat{\der_4\Xi^5}
 -\hat{S}_{\rmE}^2\widehat{\der_5\Xi^4}}L_{\rm V}^{(3)}. 
%
\eea
From (\ref{dlt_Xi:cX_T}) and (\ref{dlt_Xi:U_E}), we obtain 
\bea
 \dlt_\Xi\brkt{\frac{\cX_{\rmT}}{\UE}}
 \eql \Re\hat{\Xi}^m\der_m\brkt{\frac{\cX_{\rmT}}{\UE}}
 +\Re\brkt{\frac{1}{\hat{S}_{\rm E}^2}\widehat{\der_4\Xi^5}
 +\hat{S}_{\rm E}^2\widehat{\der_5\Xi^4}
 +i\der_m U^n\widehat{\der_n\Xi^m}}\frac{\cX_{\rmT}}{\UE}.  \label{dlt_Xi:cX_TU_E}
\eea
Therefore, we find that
\bea
 \dlt_\Xi\brkt{\frac{\cX_{\rmT}}{\UE}L_{\rm V}^{(3)}} 
 \eql \der_m\brkt{\Re\hat{\Xi}^m\frac{\cX_{\rmT}}{\UE}L_{\rm V}^{(3)}}. 
\eea

Finally, consider the last line of $\cL_{\rm VT}$. 
Combining (\ref{dlt_Xi:hatchiral}), (\ref{dlt_Xi:J_cP}), 
(\ref{dlt_Xi:J_SSbS}) and (\ref{dlt_Xi:cX_TU_E}), we can see that
\bea
 \dlt_\Xi\brkt{J_\cP f_{IJ}\frac{\cX_{\rmT}}{\UE}J_S^{(1)}\frac{\hat{S}_{\rm E}}{\bar{\hat{S}}_{\rm E}}
 \hat{\Sgm}^I\hat{\Sgm}^J} 
 \eql \der_m\brkt{\Re\hat{\Xi}^m
 J_\cP f_{IJ}\frac{\cX_{\rmT}}{\UE}J_S^{(1)}\frac{\hat{S}_{\rm E}}{\bar{\hat{S}}_{\rm E}}
 \hat{\Sgm}^I\hat{\Sgm}^J}. 
\eea
We have used the property~(\ref{prop:linear}), which also ensures that
\be
 J_\cP f_{IJ}\frac{\cX_{\rmT}}{\UE}J_S^{(1)}\frac{\hat{S}_{\rm E}}{\bar{\hat{S}}_{\rm E}}
 \hat{\Sgm}^I\hat{\Sgm}^J 
 = J_\cP f_{IJ}\frac{\cX_{\rm T}}{\UE}J_S^{(2)}
 \frac{\bar{\hat{S}}_{\rm E}}{\hat{S}_{\rm E}}\hat{\Sgm}^I\hat{\Sgm}^J. 
 \label{rel:Sgm^2}
\ee

Using the results obtained in this section, 
we can modify the action in (\ref{prev:S}) so that it is $\dlt_\Xi$-invariant 
up to total derivatives. 
We will provide the modified Lagrangian in Sec.~\ref{result}.

\section{Covariantization of $\bdm{\der_{\rmE}}$} \label{cov_derE}
So far, we have concentrated on the $\dlt_\Xi$-transformation, \ie, 
a diffeomorphism in the extra dimensions. 
In this section, we argue the consistency with its 4D counterpart, \ie, 
the 4D $\cN=1$ superconformal transformation.  
Notice that $\der_m$ does not preserve the proper transformation laws 
for the $\cN=1$ superconformal transformation collected in Appendix~\ref{N1sf_trf}. 
Thus we need to introduce the connection superfields~$\Psi_m^\alp$ 
that transform as $\dlt_L\Psi_m^\alp=-\der_m L^\alp$ 
($L^\alp$ is the $\cN=1$ superconformal transformation parameter), 
and covariantize $\der_m$.

\subsection{Chiral superspace}
On a chiral superfield, we define the covariant derivative~$\nabla_m$ as
\be
 \nabla_m \equiv \der_m-\brkt{\frac{1}{4}\bar{D}^2\Psi_m^\alp D_\alp
 -i\sgm^{\udl{\mu}}_{\alp\dalp}\bar{D}^{\dalp}\Psi_m^\alp\der_\mu
 +\frac{w}{12}\bar{D}^2D^\alp\Psi_{m\alp}}, 
 \label{cD_m:naive}
\ee
where $w$ is the Weyl weight. 
Then, $\nabla_m H$ ($H=H_{\rm odd},H_{\rm even}$) transforms as 
\bea
 \dlt_L\brkt{\nabla_m H} \eql \brkt{-\frac{1}{4}\bar{D}^2L^\alp D_\alp
 +i\sgm_{\alp\dalp}^{\udl{\mu}}\bar{D}^{\dalp}L^\alp\der_\mu
 -\frac{1}{8}\bar{D}^2D^\alp L_\alp}\nabla_mH, 
\eea
at the leading order in $\Psi^\alp$.\footnote{
In this paper, we consider the superconformal transformations at the linearized order in $\Psi_m^\alp$. 
} 
This is the same law as $\dlt_L H$. (See (\ref{dlt_L}).)
Hence, (\ref{def:L_H1}) is modified as
\be
 L_{\rm H}^{(1)} = H_{\rm odd}^t\tl{d}\brkt{\nabla_{\rmE}-\Sgm}H_{\rm even}
 -H_{\rm even}^t\tl{d}\brkt{\nabla_{\rmE}+\Sgm}H_{\rm odd}, 
 \label{cL_H:2}
\ee
where 
\be
 \nabla_{\rmE} \equiv \frac{1}{\SE}\nabla_4-\SE\nabla_5. 
\ee
This is invariant under the $\dlt_L$-transformation up to total derivatives. 

Next we consider the $\dlt_\Xi$-transformation. 
This should commute with the $\dlt_L$-transformation 
in order for the chiral property of the $\cN=1$ chiral superfields 
to be preserved. 
From this requirement, the $\dlt_\Xi$-transformation of $\Psi_m^\alp$ is found to be
\be
 \dlt_\Xi\Psi_m^\alp = \Xi^n\brkt{\der_n\Psi_m^\alp-\der_m\Psi_n^\alp}. 
 \label{dlt_Xi:Psi_m^alp}
\ee
In fact, we can see that
\be
 \dlt_L\dlt_\Xi\Psi_m^\alp = \dlt_\Xi\dlt_L\Psi_m^\alp = 0. 
\ee
The transformation law~(\ref{dlt_Xi:Psi_m^alp}) is consistent 
with the component field transformation~(\ref{dlt_xi:e_m^umu}) 
under the constraint~$\der_m\xi^{\udl{\mu}}=0$ 
if we identify the $\bar{\tht}$-component of $\Psi_{m\alp}$ as
\be
 \Psi_{m\alp} = \frac{i}{2}\brkt{\sgm^{\udl{\mu}}\bar{\tht}}_\alp 
 e_{m\udl{\mu}}+\cdots. 
\ee

Then, $\nabla_m H$ transforms as
\bea
 \dlt_\Xi\brkt{\nabla_m H} \eql \nabla_m\brkt{\dlt_\Xi H}
 -\frac{1}{4}\bar{D}^2\brkt{\dlt_\Xi\Psi_m^\alp D_\alp H}
 -\frac{1}{8}\brkt{\bar{D}^2D^\alp\dlt_\Xi\Psi_{m\alp}}H \nonumber\\
 \eql \Xi^n\nabla_n\brkt{\nabla_m H}+\nabla_m\Xi^n\nabla_n H+\nabla_m\brkt{X_\Xi H}, 
\eea
where
\be
 X_\Xi \equiv \frac{1}{4}\brkt{\nabla_m\Xi^m+\frac{1}{\SE^2}\nabla_4\Xi^5+\SE^2\nabla_5\Xi^4}. 
\ee
We have used that
\bea
 \nabla_m\brkt{\nabla_n H} \eql \nabla_n\brkt{\nabla_m H}
 -\frac{1}{4}\bar{D}^2\brc{\brkt{\der_m\Psi_n^\alp-\der_n\Psi_m^\alp}D_\alp H} \nonumber\\
 &&-\frac{1}{8}\brc{\bar{D}^2D^\alp\brkt{\der_m\Psi_{n\alp}-\der_n\Psi_{m\alp}}}H+\cO(\Psi^2). 
\eea
As a result, the $\dlt_\Xi$-transformation of (\ref{cL_H:2}) becomes total derivatives. 
\bea
 \dlt_\Xi L_{\rm H}^{(1)} \eql \nabla_m\brkt{\Xi^m L_{\rm H}^{(1)}} \nonumber\\
 \eql \der_m\brkt{\Xi^m L_{\rm H}^{(1)}}
 -\frac{1}{4}\bar{D}^2D^\alp\brkt{\Psi_{m\alp}\Xi^m L_{\rm H}^{(1)}}. 
\eea
Note that $L_{\rm H}^{(1)}$ has the Weyl weight~3.

\subsection{Full superspace}
In the full superspace, $\nabla_m$ in (\ref{cD_m:naive}) is modified as
\bea
 \tl{\nabla}_m \defa \der_m-\brkt{\frac{1}{4}\bar{D}^2\Psi_m^\alp D_\alp
 +\frac{1}{2}\bar{D}^{\dalp}\Psi_m^\alp\bar{D}_{\dalp}D_\alp
 +\frac{w+n}{24}\bar{D}^2\cR_U D^\alp\cR_U^{-1}\Psi_{m\alp}} \nonumber\\
 &&-\cR_U\brkt{\frac{1}{4}D^2\bar{\Psi}_{m\dalp}\bar{D}^{\dalp}
 +\frac{1}{2}D^\alp\bar{\Psi}_m^{\dalp}D_\alp\bar{D}_{\dalp}
 +\frac{w-n}{24}D^2\cR_U^{-1}\bar{D}_{\dalp}\cR_U\bar{\Psi}_m^{\dalp}}, 
\eea
where $n$ is the chiral weight (\ie, the $U(1)_A$ charge), 
and the operator~$\cR_U$ is defined by
\be
 \cR_U X_m = X_m-2iU^n\brkt{\der_n X_m-\der_m X_n}+\cO(U^2). 
\ee
Then, from the relation: 
\bea
 \tl{\nabla}_m\tl{\nabla}_n \eql \tl{\nabla}_n\tl{\nabla}_m
 -\left\{\frac{1}{4}\bar{D}^2\brkt{\der_m\Psi_n^\alp-\der_n\Psi_m^\alp}D_\alp
 +\frac{1}{2}\bar{D}^{\dalp}\brkt{\der_m\Psi_n^\alp-\der_n\Psi_m^\alp}\bar{D}_{\dalp}D_\alp 
 \right.\nonumber\\
 &&\hspace{20mm}\left.
 +\frac{w+n}{24}\bar{D}^2D^\alp\brkt{\der_m\Psi_{n\alp}-\der_n\Psi_{m\alp}}+\hc\right\}
 +\cO(U^m), 
\eea
and the transformation law: 
\be
 \dlt_\Xi\brkt{\cP_U^{-1}V^I} = \Xi^n\tl{\nabla}_n\brkt{\cP_U^{-1}V^I}, 
\ee
we find that
\be
 \dlt_\Xi\brc{\tl{\nabla}_m\brkt{\cP_U^{-1}V^I}} 
 = \Xi^n\tl{\nabla}_n\brc{\tl{\nabla}_m\brkt{\cP_U^{-1}V^I}}
 +\tl{\nabla}_m\Xi^n\tl{\nabla}_n\brkt{\cP_U^{-1}V^I}+\cO(U^m), 
\ee
which leads to 
\bea
 \dlt_\Xi\brc{\tl{\nabla}_{\rm E}\brkt{\cP_U^{-1}V}}
 \eql \Xi^n\tl{\nabla}_n\brc{\tl{\nabla}_{\rm E}\brkt{\cP_U^{-1}V^I}} \nonumber\\
 &&+\frac{1}{2}\brkt{\tl{\nabla}_n\Xi^n-\frac{1}{\SE^2}\tl{\nabla}_4\Xi^5
 -\SE^2\tl{\nabla}_5\Xi^4}\tl{\nabla}_{\rm E}\brkt{\cP_U^{-1}V^I} 
 +\cO(U^m), \nonumber\\
 \label{dlt_Xi:der_EV:2}
\eea
where
\be
 \tl{\nabla}_{\rm E} \equiv \frac{1}{\SE}\tl{\nabla}_4-\SE\tl{\nabla}_5. 
\ee

Therefore, $L_{\rm V}^{(1)\alp}$ in (\ref{def:L_V^1}) and $L_{\rm V}^{(3)}$ in (\ref{def:L_V^3}) 
are modified as
\bea
 L_{\rm V}^{(1)\alp} \eql f_{IJ}\left\{-2\hat{\Sgm}^I D^{\cP\alp}V^J
 +\frac{1}{2}\brkt{\nabla_{\rm E}^\cP V^I D^{\cP\alp}V^J
 -\nabla_{\rm E}^\cP D^{\cP\alp}V^IV^J}\right\}, \nonumber\\
 L_{\rm V}^{(3)} \eql f_{IJ}\left\{4\brkt{\nabla_{\rm E}^\cP V^I-\hat{\Sgm}^I}^\dagger
 \brkt{\nabla_{\rm E}^\cP V^J-\hat{\Sgm}^J}
 -2\brkt{\nabla_{\rm E}^\cP V^I}^\dagger \nabla_{\rm E}^\cP V^J\right\}, 
\eea
where 
\be
 \nabla_{\rm E}^\cP \equiv \cP_U\tl{\nabla}_{\rm E}\cP_U^{-1}. 
\ee

Besides, the $\dlt_\Xi$-transformations~(\ref{dlt_Xi:hatchiral}), (\ref{dlt_Xi:V^I}) 
and (\ref{dlt_Xi:V_Tm}) are modified as
\bea
 \dlt_\Xi\hat{S}_{\rm E} \eql (\Re\hat{\Xi}^m)\nabla_m^\cP\hat{S}_{\rm E}
 +\frac{1}{2}\brkt{\widehat{\nabla_4\Xi^4}-\widehat{\nabla_5\Xi^5}
 +\frac{1}{\hat{S}_{\rm E}^2}\widehat{\nabla_4\Xi^5}
 -\hat{S}_{\rm E}^2\widehat{\nabla_5\Xi^4}}\hat{S}_{\rm E}, \nonumber\\
 \dlt_\Xi\hat{H} \eql (\Re\hat{\Xi}^m)\nabla_m^\cP\hat{H}
 +\frac{1}{4}\brkt{\widehat{\nabla_m\Xi^m}
 +\frac{1}{\hat{S}_{\rm E}^2}\widehat{\nabla_4\Xi^5}
 +\hat{S}_{\rm E}^2\widehat{\nabla_5\Xi^4}}\hat{H}, \nonumber\\
 \dlt_\Xi\hat{\Sgm}^I \eql (\Re\hat{\Xi}^m)\nabla_m^\cP\hat{\Sgm}^I
 +\frac{1}{2}\brkt{\widehat{\nabla_m\Xi^m}
 -\frac{1}{\hat{S}_{\rm E}^2}\widehat{\nabla_4\Xi^5}
 -\hat{S}_{\rm E}^2\widehat{\nabla_5\Xi^4}}\hat{\Sgm}^I, \nonumber\\
 \dlt_\Xi V^I \eql (\Re\hat{\Xi}^m)\Re\brkt{\nabla_m^\cP V^I}, \nonumber\\
 \dlt_\Xi\Ups_{\rmT\alp} \eql \Xi^m\nabla_m\Ups_{\rmT\alp}, \nonumber\\
 \dlt_\Xi V_{\rmT m} \eql (\Re\hat{\Xi}^n)\Re\brkt{\nabla_n^\cP V_{\rmT m}}
 +(\Re\nabla_m^\cP\hat{\Xi}^n)V_{\rmT n}, \nonumber\\
 \dlt_\Xi\Sgm_{\rmT} \eql \nabla_m\brkt{\Xi^m\Sgm_{\rmT}}, 
\eea
where
\be
 \nabla_m^\cP \equiv \cP_U\tl{\nabla}_m\cP_U^{-1}, \;\;\;\;\;
 \widehat{\nabla_m\Xi^n} \equiv \cP_U\brkt{\nabla_m\Xi^n}. 
\ee

\section{Rotations that mix 4D and extra dimensions} \label{Lorentz_inv}
Here we consider the Lorentz transformations that mix 4D and the extra dimensions. 
In order to simplify the discussion, we treat the ``off-diagonal'' superfields~$U^m$ and $\Psi_m^\alp$ 
at the linearized level in this section.  
Then, the corresponding superfield transformation laws are given by
\bea
 \dlt_N U^\mu \eql 0, \;\;\;\;\;
 \dlt_N U^4 = \Re\brkt{\frac{N}{\SE}}, \;\;\;\;\;
 \dlt_N U^5 = -\Re\brkt{N\SE}, \nonumber\\
 \dlt_N\tl{V}_{\rm E} \eql 2\tl{V}_{\rm E}^{1/2}\Im\derE\brkt{N\tl{V}_{\rm E}^{1/2}}, \;\;\;\;\;
 \dlt_N\SE = 0, \nonumber\\
 \dlt_N\Psi_4^\alp \eql -\frac{iV_{\rm E}}{2}
 D^\alp\Im\brkt{N\SE}, \;\;\;\;\;
 \dlt_N\Psi_5^\alp = -\frac{iV_E}{2}D^\alp\Im\brkt{\frac{N}{\SE}},  \nonumber\\
 \dlt_N H_{\rm odd} \eql -\frac{i}{4}\bar{D}^2\brkt{N\tl{V}_{\rm E}^{1/2}e^{-V}\bar{H}_{\rm even}}, 
 \;\;\;\;\;
 \dlt_N H_{\rm even} = \frac{i}{4}\bar{D}^2\brkt{N\tl{V}_{\rm E}^{1/2}e^V\bar{H}_{\rm odd}}, 
 \nonumber\\
 \dlt_N V^I \eql \Im\brc{N\brkt{\derE V^I-2\Sgm^I}}, \;\;\;\;\;
 \dlt_N\Sgm^I = -\frac{i}{8}\bar{D}^2\brkt{\tl{V}_{\rm E}D^\alp\bar{N} D_\alp V^I}, 
 \label{dlt_Ns}
\eea
where $\tl{V}_{\rm E}\equiv\VE \UE$, 
and the transformation parameter~$N$ is a complex general superfield 
whose $\tht\bar{\tht}$-component is 
\bea
 N \eql (\tht\sgm^{\udl{\mu}}\bar{\tht})\brkt{\lmd_{\udl{\mu}\udl{4}}-i\lmd_{\udl{\mu}\udl{5}}}
 \frac{\sqrt{E_4E_5}}{ie^{(2)}}+\cdots. 
\eea

\subsection{Invariance in hyper sector}
The invariance of the action under the $\dlt_N$-transformation is less manifest than the $\dlt_L$- 
and the $\dlt_\Xi$-transformations because the cancellation between the $\int\dr^4\tht$- 
and the $\int\dr^2\tht$-integrals occurs in the $\dlt_N$-transformation. 
Here, we show the invariance in the hyper sector to illustrate such cancellation. 

From (\ref{dlt_Ns}), the hatted superfields transform as 
\bea
 \dlt_N\hat{H}_{\rm odd} \eql \frac{i}{2}\brkt{N\derE+\bar{N}\bar{\der}_{\rm E}}H_{\rm odd}
 -\frac{i}{4}\bar{D}^2\brkt{N\tl{V}_{\rm E}^{1/2}e^{-V}\bar{H}_{\rm even}}, \nonumber\\
 \dlt_N\hat{H}_{\rm even} \eql \frac{i}{2}\brkt{N\derE+\bar{N}\bar{\der}_{\rm E}}H_{\rm even}
 +\frac{i}{4}\bar{D}^2\brkt{N\tl{V}_{\rm E}^{1/2}e^V\bar{H}_{\rm odd}}. 
\eea
After some straightforward calculations, we can see that 
$L_{\rm H}^{(1)}$ in (\ref{cL_H:2}) and $L_{\rm H}^{(2)}$ in (\ref{def:L_H^2}) transform as
\bea
 \dlt_N L_{\rm H}^{(1)} \eql -\frac{i}{4}\bar{D}^2\left[
 2N\tl{V}_{\rm E}^{1/2}\brc{
 H_{\rm odd}^\dagger\tl{d}\brkt{\derE+\Sgm}H_{\rm odd} 
 +H_{\rm even}^\dagger\tl{d}\brkt{\derE-\Sgm}H_{\rm even}}
 \right. \nonumber\\
 &&\hspace{13mm}
 +N\tl{V}_{\rm E}^{1/2}(\cO_{\rm E}\SE)\brkt{H_{\rm odd}^\dagger\tl{d} H_{\rm odd}
 +H_{\rm even}^\dagger\tl{d} H_{\rm even}} \nonumber\\
 &&\hspace{13mm}\left. 
 +\frac{1}{2}\tl{V}_{\rm E}D^\alp\bar{N}\brkt{H_{\rm odd}^t\tl{d} D_\alp H_{\rm even}
 -H_{\rm even}^t\tl{d} D_\alp H_{\rm odd}-2D_\alp V H_{\rm odd}^t\tl{d}H_{\rm even}}\right], 
 \nonumber\\
\eea
and
\bea
 \dlt_N\brkt{-2\tl{\VE}^{1/2}L_{\rm H}^{(2)}} 
 \eql \Im\left[4N\tl{V}_{\rm E}^{1/2}
 \brc{H_{\rm odd}^\dagger\tl{d}e^V\brkt{\derE+\Sgm}H_{\rm odd}
 +H_{\rm even}^\dagger\tl{d}e^{-V}\brkt{\derE-\Sgm}H_{\rm even}} \right.\nonumber\\
 &&\hspace{7mm}
 -2N\tl{V}_{\rm E}^{1/2}(\cO_{\rm E}\SE)
 \brkt{H_{\rm odd}^\dagger\tl{d}e^V H_{\rm odd}+H_{\rm even}^\dagger\tl{d}e^{-V}H_{\rm even}} \nonumber\\
 &&\hspace{7mm}
 +\tl{V}_{\rm E}^{1/2}\left\{
 H_{\rm odd}^t\tl{d}e^VD^2\brkt{\bar{N}\tl{V}_{\rm E}^{1/2}e^{-V}H_{\rm even}} \right.\nonumber\\
 &&\hspace{20mm}\left.\left.
 -H_{\rm even}^t\tl{d}e^{-V}D^2\brkt{\bar{N}\tl{V}_{\rm E}^{1/2}e^VH_{\rm odd}}
 \right\}\right], 
\eea
up to total derivatives. 
We have dropped the $U^m$- and the $\Psi_m^\alp$-dependent terms in the right-hand-sides. 
The last line in $\dlt_N L_{\rm H}^{(2)}$ can be rewritten as
\bea
 A \defa \tl{V}_{\rm E}^{1/2}\brc{H_{\rm odd}^t\tl{d}e^V
 D^2\brkt{\bar{N}\tl{V}_{\rm E}^{1/2}e^{-V}H_{\rm even}}
 -H_{\rm even}^t\tl{d}e^{-V}D^2\brkt{\bar{N}\tl{V}_{\rm E}^{1/2}e^VH_{\rm odd}}} \nonumber\\
 \eql 2D^\alp\bar{N}\brkt{\tl{H}_{\rm o}^t\tl{d}D_\alp\tl{H}_{\rm e}-\tl{H}_{\rm e}^t\tl{d}D_\alp\tl{H}_{\rm o}}
 +\bar{N}\brkt{\tl{H}_{\rm o}^t\tl{d}D^2\tl{H}_{\rm e}-\tl{H}_{\rm e}^t\tl{d}D^2\tl{H}_{\rm o}}, 
\eea
where
\be
 \tl{H}_{\rm o} \equiv \tl{V}_{\rm E}^{1/2}e^V H_{\rm odd}, \;\;\;\;\;
 \tl{H}_{\rm e} \equiv \tl{V}_{\rm E}^{1/2}e^{-V} H_{\rm even}. 
\ee
This can also be rewritten as
\be
 A = \bar{N}\brkt{\tl{H}_{\rm e}^t\tl{d}D^2\tl{H}_{\rm o}-\tl{H}_{\rm o}^t\tl{d}D^2\tl{H}_{\rm e}}, 
\ee
up to total derivatives. 
Therefore, we obtain 
\bea
 A \eql D^\alp\bar{N}\brkt{\tl{H}_{\rm o}\tl{d}D_\alp\tl{H}_{\rm e}
 -\tl{H}_{\rm e}^t\tl{d}D_\alp\tl{H}_{\rm o}} \nonumber\\
 \eql \tl{V}_{\rm E}D^\alp\bar{N}
 \brkt{H_{\rm odd}^t\tl{d}D_\alp H_{\rm even}
 -H_{\rm even}^t\tl{d}D_\alp H_{\rm odd}
 -2D_\alp V H_{\rm odd}^t\tl{d}H_{\rm even}}. 
\eea
We should also note that 
\bea
 \dlt_N\abs{J_\cP} \eql \cO(U^m), 
\eea
since $\abs{J_\cP}=1+\cO((U^m)^2)$. 

Making use of these, we can show that
\be
 \dlt_N\cL_{\rm H} = \dlt_N\brc{-2\int\dr^4\tht\;\abs{J_\cP}\tl{\VE}^{1/2}L_{\rm H}^{(2)}
 +\brkt{\int\dr^2\tht\;L_{\rm H}^{(1)}+\hc}} = 0, 
\ee
up to total derivatives. 
We have used the relation~$\int\dr^2\bar{\tht}=-\frac{1}{4}\bar{D}^2$ 
in the $d^2\tht$-integration.

\subsection{Kinetic terms for $\bdm{U^m}$ and $\bdm{\Psi_m^\alp}$}
Now we consider the kinetic terms for the gravitational superfields, 
which originate from the 6D Weyl multiplet. 
Among $\brc{U^\mu,U^m,\Psi_m^\alp,\VE,\SE}$, 
only $\VE$ and $\SE$ have nonvanishing background values. 
Here, we treat the superfields~$\brc{U^\mu,U^m,\Psi_m^\alp}$ 
and the fluctuation parts of $\VE$ and $\SE$ at the linearized order, 
and neglect terms beyond quadratic in them. 
As shown in Appendix~\ref{4DSUGRAcouplings}, the kinetic term for $U^\mu$, $\cL_{\rm E}^{N=1}$,  
is given by (\ref{kin_for_U^mu}). 
There is an additional term that involves the ``off-diagonal'' component superfields~$U^m$ 
and $\Psi_m^\alp$.  

We define the covariant derivatives of $U^\mu$ as
\be
 \tl{\nabla}_m U^\mu \equiv \der_m U^\mu-\frac{1}{2}\sgm_{\alp\dalp}^\mu
 \brkt{\bar{D}^{\dalp}\Psi_m^\alp-D^\alp\bar{\Psi}_m^{\dalp}}, 
\ee
where $\sgm_{\alp\dalp}^\mu = \vev{e_{\udl{\nu}}^{\;\;\mu}}\sgm^{\udl{\nu}}_{\alp\dalp}$. 
This has the Weyl weight 0, and is invariant under the $\dlt_L$-transformation. 
In order to construct the $\dlt_N$-invariant term, 
we redefine the above covariant derivatives as 
\bea
 \nabla_4 U^\mu \defa \tl{\nabla}_4 U^\mu
 +\VE\brc{\brkt{\Im\SE^2}\der^\mu U^4-\frac{i\Re\SE^2}{\bar{S}_{\rm E}^2}\der^\mu U^5}, \nonumber\\
 \nabla_5 U^\mu \defa \tl{\nabla}_5 U^\mu
 +\VE\brc{\frac{\Im\SE^2}{\abs{\SE}^4}\der^\mu U^5
 +\frac{i\Re\SE^2}{\SE^2}\der^\mu U^4}, 
\eea
where $\der^\mu \equiv \vev{e_{\udl{\rho}}^{\;\;\mu}}\vev{e_{\udl{\tau}}^{\;\;\nu}}\eta^{\udl{\rho}\udl{\tau}}
\der_\nu$. 
Then, the combination:
\bea 
 \cC_{\rm E}^\mu \defa \frac{1}{\SE}\nabla_4 U^\mu-\SE\nabla_5 U^\mu \nonumber\\
 \eql \frac{1}{\SE}\tl{\nabla}_4 U^\mu-\SE\tl{\nabla}_5 U^\mu
 -i\VE\brkt{\SE\der^\mu U^4+\frac{\der^\mu U^5}{\SE}}  \label{def:cC_E}
\eea
is $\dlt_L$- and $\dlt_N$-invariant at the linearized order. 
\be
 \dlt_L\cC_{\rm E}^\mu = \cO(U^m), \;\;\;\;\;
 \dlt_N\cC_{\rm E}^\mu = \cO(U^m,D^\alp\VE,D^\alp\SE). 
\ee
Using this combination, we can construct 
the following $\dlt_L$- and $\dlt_N$-invariant Lagrangian term. 
\be
 \cL_{\cC} = \int\dr^4\tht\;a\bar{\cC}_{\rm E}^\mu\cC_{{\rm E}\mu}, 
\ee
where $a$ is a real constant. 
The invariance of the action under the $\dlt_\Xi$-transformation determines $a$. 
Restoring the $U^\mu$-dependence (see Appendix~\ref{4DInv_action}), 
the 6D Lagrangian should have the form of
\be
 \cL = \cL_{\rm E}^{N=1}+\cL_{\cC}+\int\dr^4\tht\;\brkt{1+\frac{1}{12}\bar{\sgm}_\mu^{\dalp\alp}
 \sbk{D_\alp,\bar{D}_{\dalp}}U^\mu}\Omg+\brkt{\int\dr^2\tht\;W+\hc}, 
\ee
where $\Omg$ and $W$ are real and holomorphic functions respectively, 
whose explicit forms will be given in Sec.~\ref{result}. 
Recall that $\dlt_\Xi\Omg=\der_m\brkt{\Re\hat{\Xi}^m\Omg}$ 
from the results in Sec.~\ref{Xi-invariance}. 
Then, we have
\bea
 \dlt_\Xi\cL \eql \dlt_\Xi\cL_{\cC}+\int\dr^4\tht\;\brkt{\frac{1}{12}\bar{\sgm}_{\udl{\mu}}^{\dalp\alp}
 \sbk{D_\alp,\bar{D}_{\dalp}}U^\mu}\der_m\brkt{\Re\Xi^m\Omg}+\cdots \nonumber\\
 \eql \dlt_\Xi\cL_{\cC}-\int\dr^4\tht\;\frac{\Omg}{12}\der_m U^\mu
 \bar{\sgm}_{\udl{\mu}}^{\dalp\alp}\sbk{D_\alp,\bar{D}_{\dalp}}\Re\Xi^m+\cdots \nonumber\\
 \eql \dlt_\Xi\cL_{\cC}-\int\dr^4\tht\;\frac{\Omg}{3}\der_m U^\mu\Im\der_\mu\Xi^m+\cdots, 
\eea
where we have dropped total derivatives, 
and also dropped the fluctuation part of $\Omg$.\footnote{
The superconformal gauge-fixing condition~$\Omg|_{\tht=0}=-3M_{\rm 6D}^4$ must be imposed 
in order to obtain the Poincar\'e SUGRA. ($M_{\rm 6D}$ is the 6D Planck mass.)
}
Here, since 
\bea
 \dlt_\Xi\cC_{\rm E}^\mu 
 \eql i\VE\brkt{\SE\Im\der^\mu\Xi^4+\frac{1}{\SE}\Im\der^\mu\Xi^5}+\cdots, 
\eea
where the ellipses are of $\cO(U^\mu,U^m,\Psi_m,D^\alp\VE,D^\alp\SE)$, 
we can see that
\bea
 \dlt_\Xi\cL_{\cC} \eql \int\dr^4\tht\;a\brc{\bar{\der}_{\rm E}U^\mu\cdot 
 i\VE\brkt{\SE\Im\der_\mu\Xi^4+\frac{1}{\SE}\Im\der_\mu\Xi^5}}+\hc+\cdots \nonumber\\
 \eql -\int\dr^4\tht\;2a\VE\Im\brkt{\frac{\SE}{\bar{S}_{\rm E}}\der_4 U^\mu\Im\der_\mu\Xi^4
 -\frac{\bar{S}_{\rm E}}{\SE}\der_5 U^\mu\Im\der_\mu\Xi^5}+\cdots \nonumber\\
 \eql \int\dr^4\tht\;2a\VE \UE\der_m U^\mu\Im\der_\mu\Xi^m+\cdots. 
\eea
Therefore, from the $\dlt_\Xi$-invariance of the action, we find 
\be
 a = \Lvev{\frac{\Omg}{6\VE \UE}}. 
\ee

\subsection{6D SUGRA Lagrangian} \label{result}
Here we summarize our results. 
The 6D SUGRA Lagrangian is expressed as
\be
 \cL = \int\dr^4\tht\;L_{\rm E}
 +\int\dr^4\tht\;\brkt{1+\frac{1}{12}\bar{\sgm}_\mu^{\dalp\alp}\sbk{D_\alp,\bar{D}_{\dalp}}U^\mu}
 \Omg_{\rm HVT}
 +\brkt{\int\dr^2\tht\;L_{\rm H}^{(1)}+\hc}, 
\ee
where
\bea
 L_{\rm E} \defa \frac{\vev{\Omg_{\rm HVT}}}{3}\brc{\frac{1}{8}U^\mu D^\alp\bar{D}^2D_\alp U_\mu
 +\frac{1}{48}\brkt{\bar{\sgm}_\mu^{\dalp\alp}\sbk{D_\alp,\bar{D}_{\dalp}}U^\mu}^2
 -\brkt{\der_\mu U^\mu}^2+\frac{\bar{\cC}_{\rm E}^\mu\cC_{{\rm E}\mu}}{2\vev{\VE \UE}}}, 
 \nonumber\\
 \cC_{\rm E}^\mu \defa \derE U^\mu-\frac{1}{2}\sgm_{\alp\dalp}^{\mu}
 \brc{\frac{1}{\SE}\brkt{\bar{D}^{\dalp}\Psi_4^\alp-D^\alp\bar{\Psi}_4^{\dalp}}
 -\SE\brkt{\bar{D}^{\dalp}\Psi_5^\alp-D^\alp\bar{\Psi}_5^{\dalp}}} \nonumber\\
 &&-i\VE\der^\mu\brkt{\SE U^4+\frac{U^5}{\SE}}, 
 \nonumber\\
 \Omg_{\rm HVT} \defa 
 -2\abs{J_\cP}\VE^{1/2}\UE^{1/2} L_{\rm H}^{(2)}
 +\brkt{J_\cP L_{\rm V}^{(1)\alp}\hat{\cY}_{{\rm T}\alp}+\hc}
 +\cV_{\rm T}L_{\rm V}^{(2)} \nonumber\\
 &&+\frac{\cX_{\rm T}}{\UE}L_{\rm V}^{(3)}
 +\brkt{J_\cP\frac{\cX_{\rmT}}{\UE}L_{\rm V}^{(4)}+\hc}, 
\eea
and
\bea
 L_{\rm H}^{(1)} \defa H^t_{\rm odd}\tl{d}\brkt{\nabla_{\rm E}-\Sgm}H_{\rm even}
 -H^t_{\rm even}\tl{d}\brkt{\nabla_{\rm E}+\Sgm}H_{\rm odd}, \nonumber\\
 L_{\rm H}^{(2)} \defa \hat{H}_{\rm odd}^\dagger\tl{d}e^V\hat{H}_{\rm odd}
 +\hat{H}_{\rm even}^\dagger\tl{d}e^{-V}\hat{H}_{\rm even}, \nonumber\\
 L_{\rm V}^{(1)\alp} \defa f_{IJ}\brc{-2\hat{\Sgm}^I D^{\cP\alp}V^J
 +\frac{1}{2}\brkt{\nabla_{\rm E}^\cP V^I D^{\cP\alp}V^J-\nabla_{\rm E}^\cP D^{\cP\alp}V^I V^J}}, 
 \nonumber\\
 L_{\rm V}^{(2)} \defa f_{IJ}\brkt{D^{\cP\alp}V^I\hat{\cW}_\alp^J
 +\frac{1}{2}V^ID^{\cP\alp}\hat{\cW}_\alp^J+\hc}, \nonumber\\
 L_{\rm V}^{(3)} \defa f_{IJ}\brc{4\brkt{\nabla_{\rm E}^\cP V^I-\hat{\Sgm}^I}^\dagger
 \brkt{\nabla_{\rm E}^\cP V^J-\hat{\Sgm}^J}
 -2\brkt{\nabla_{\rm E}^\cP V^I}^\dagger\nabla_{\rm E}^\cP V^J}, \nonumber\\
 L_{\rm V}^{(4)} \defa f_{IJ}J_S^{(1)}\frac{2\hat{S}_{\rm E}}{\bar{\hat{S}}_{\rm E}}
 \hat{\Sgm}^I\hat{\Sgm}^J. 
 \label{6DSUGRA:Ls}
\eea
The covariant derivatives~$\nabla_{\rm E}$, $\nabla_{\rm E}^\cP$ 
and $\nabla_m^\cP$ are defined in Sec.~\ref{cov_derE}, 
and the field strengths are given by 
\bea
 \hat{\cW}_\alp^I \defa -\frac{1}{4}(\bar{D}^\cP)^2D_\alp^\cP V^I+\cO(U^\mu,U^m\Sgm), 
 \nonumber\\
 \cX_{\rmT} \defa \frac{1}{2}\Im\brkt{D^{\cP\alp}\Ups_{\rmT\alp}}, \nonumber\\
 \cY_{{\rm T}\alp} \defa \frac{1}{2\SE}\cW_{{\rm T}4\alp}+\frac{\SE}{2}\cW_{{\rm T}5\alp}
 +\frac{1}{2}\brkt{\frac{1}{\SE}\nabla_4+\SE\nabla_5}\Ups_{{\rm T}\alp}, \nonumber\\
 \cV_{\rm T} \defa \Re\brkt{\nabla_{\rm 4}^\cP V_{\rm T 5}-\nabla_5^\cP V_{\rm T4}}
 +J_\cP\hat{\Sgm}_{\rm T}+\bar{J}_\cP\bar{\hat{\Sgm}}_{\rm T}, 
 \label{6DSUGRAfct}
\eea
and $J_\cP$, $\UE$ and $J_S^{(1)}$ are defined in (\ref{def:J_cP}), 
(\ref{def:U_E}) and (\ref{def:J_S}), respectively. 

We have revived the $U^\mu$-dependence. 
Thus, for a chiral superfield~$\Phi$, $\hat{\Phi}$ should be understood as
\be
 \hat{\Phi}(x^M,\tht,\bar{\tht}) = \Phi(x^M+iU^M(\tht,\bar{\tht}),\tht,\bar{\tht}). 
\ee
The $U^\mu$-dependence of $\hat{\cW}_\alp^I$ is given by (\ref{def:E_1}). 

The real superfield~$\VE$ is expressed as
\be
 \VE = \frac{\cV_{\rmT}}{\cX_{\rmT}},  \label{sol:V_E}
\ee
and the chiral superfield~$\Ups_{{\rm T}\alp}$ is subject to the constraint:
\be
 \frac{1}{\SE}\cW_{{\rm T}4\alp}-\SE\cW_{{\rm T}5\alp}+\nabla_{\rm E}\Ups_{{\rm T}\alp} = 0. 
 \label{constraint_on_cY}
\ee
Note that this contains $\Psi_m^\alp$ ($m=4,5$). 
This constraint indicates that either $\Psi_4^\alp$ or $\Psi_5^\alp$ is a dependent superfield, 
\ie, it can be expressed in terms of the other superfields.

\section{Dimensional reduction to 5D} \label{dim_red:5D}
We consider the situation that the two extra dimensions are compactified on a torus, 
\ie, $x^m\in\sbk{0,L_m}$. 
We take the coordinates so that $L_m=\cO(1)$. 
Since the ``off-diagonal'' components of the sechsbein do not have nonvanishing background values, 
the line element along the extra dimensions is expressed as
\be
 ds^2 = \vev{e_m^{\;\;\udl{4}}e_n^{\;\;\udl{4}}+e_m^{\;\;\udl{5}}e_n^{\;\;\udl{5}}}dx^m dx^n 
 = \abs{\vev{E_m} dx^m}^2. 
\ee
Hence, the ratio of the sizes of the two extra dimensions is parameterized by 
the background value of $\SE$ because 
\be
 \abs{\vev{\SE}}^2 = \frac{\abs{\vev{E_4}}}{\abs{\vev{E_5}}}. 
\ee
Therefore, the limit that the sixth (fifth) dimension shrinks to zero corresponds to 
the limit~$\abs{\SE}\to\infty$ ($\abs{\SE}\to 0$). 
Since the extra dimensions are compactified, there are mass gaps between the zero-modes 
and the KK excited modes. 
For the latter, $\der_m$ gives $\cO(1)$ factors because we have taken $L_m$ as $\cO(1)$. 
When $\abs{\SE}\to \infty$ ($\abs{\SE}\to 0$), 
terms involving $\der_5$ ($\der_4$) in $\nabla_{\rm E}$ grow infinitely large 
and drop out of the path integral. 
So we can neglect such terms 
because only the contributions from the zero-modes survive. 
In such a case, we should drop the covariant derivative~$\nabla_5$ ($\nabla_4$) 
in order to maintain the 4D diffeomorphism invariance. 
As a result, we can replace $\nabla_E$ with $\frac{1}{\SE}\nabla_4$ ($-\SE\nabla_5$) 
in this limit. 

Let us consider the limit~$\abs{\SE}\to\infty$ as an example.\footnote{
The procedure in the limit~$\abs{\SE}\to 0$ is similar if we use the relation~(\ref{rel:Sgm^2}). 
}
In this case, we can neglect the $x^5$-dependence of the superfields, 
and the only extra-dimensional coordinate is $y\equiv x^4$. 
Thus $\cP_U$ is understood as the operator that shifts $y$ as 
$y\to y+iU^4$.

\subsection{Hyper sector}
First, we consider the hyper sector. 
The covariant derivative~$\nabla_{\rm E}$ becomes 
\bea
 \nabla_{\rm E} \toa 
 \frac{1}{\SE}\nabla_y^{\rm (5D)}, \nonumber\\
 \nabla_y^{\rm (5D)} \defa 
 \der_y-\frac{1}{4}\bar{D}^2\brkt{\Psi_y^\alp D_\alp}-\frac{w}{12}\bar{D}^2D^\alp\Psi_{y\alp},  
 \label{limit:nabla_E}
\eea
where $\Psi_y^\alp\equiv\Psi_4^\alp$. 
Thus, $L_{\rm H}^{(1)}$ in (\ref{6DSUGRA:Ls}) becomes 
\be
 L_{\rm H}^{(1)} \to 
 H_{\rm odd}^{{\rm (5D)}t}\tl{d}\brkt{\nabla_y^{\rm (5D)}-\Sgm^{\rm (5D)}}
 H_{\rm even}^{\rm (5D)}
 -H_{\rm even}^{{\rm (5D)}t}\tl{d}\brkt{\nabla_y^{\rm (5D)}+\Sgm^{\rm (5D)}}H_{\rm odd}^{\rm (5D)}, 
 \label{L_H^5D:1}
\ee
where 
\be
 H_{\rm odd}^{\rm (5D)} \equiv \SE^{-1/2}H_{\rm odd}, \;\;\;\;\;
 H_{\rm even}^{\rm (5D)} \equiv \SE^{-1/2}H_{\rm even}, \;\;\;\;\;
 \Sgm^{{\rm (5D)}I} \equiv \SE\Sgm^I. 
\ee

As for the full superspace part, we obtain 
\be
 \abs{J_\cP}\VE^{1/2}\UE^{1/2} L_{\rm H}^{(2)} \to 
 \abs{J_y}V_{\rm E}^{\rm (5D)1/2}\brkt{
 \hat{H}_{\rm odd}^{\rm (5D)\dagger}\tl{d}e^V\hat{H}_{\rm odd}^{\rm (5D)}
 +\hat{H}_{\rm even}^{\rm (5D)\dagger}\tl{d}e^{-V}\hat{H}_{\rm even}^{\rm (5D)}}, 
 \label{L_H^5D:2}
\ee
where
\be
 J_y \equiv 1+i\der_y U^4, \;\;\;\;\;
 V_{\rm E}^{\rm (5D)} \equiv \VE\UE|\hat{S}_{\rm E}|^2. 
 \label{def:V_E^5D}
\ee

The integrands~(\ref{L_H^5D:1}) and (\ref{L_H^5D:2}) agree with 
those in Ref.~\cite{Sakamura:2012bj} at the linearized order in $U^4$.

\subsection{Vector-tensor sector}
Next consider the vector-tensor sector. 
Noting that 
\bea
 \der_4^\cP V^I \eql \cP_U\der_4\cP_U^{-1}V^I \nonumber\\
 \eql \der_4 V^I-i\der_4 U^m\der_m V^I+(-i)^2\der_4 U^m\der_m U^n\der_n V^I+\cdots \nonumber\\
 \toa \sum_{n=0}^\infty\brkt{-i\der_4 U^4}^n\der_4 V^I 
 = \frac{1}{1+i\der_4 U^4}\der_4 V^I 
 = \frac{\der_4 V^I}{J_y}, 
\eea
the covariant derivative~$\nabla_{\rm E}^\cP$ becomes
\be
 \nabla_{\rm E}^\cP \to \frac{1}{J_y\hat{S}_{\rm E}}\nabla_y^{{\rm (5D)}\cP}
 +\cO(\Psi_y U^4), 
 \label{limit:nabla_E^cP}
\ee
where
\be
 \nabla_y^{{\rm (5D)}\cP} \equiv \der_y-\brkt{\frac{1}{4}\bar{D}^2\Psi_y^\alp D_\alp
 +\frac{1}{2}\bar{D}^{\dalp}\Psi_y^\alp\bar{D}_{\dalp}D_\alp
 +\frac{w+n}{24}\bar{D}^2D^\alp\Psi_{y\alp}+\hc}+\cO(\Psi_y U^4). 
\ee
Therefore, we obtain
\be
 L_{\rm V}^{(1)\alp} \to \frac{f_{IJ}}{\hat{S}_{\rm E}}\brc{-2\hat{\Sgm}^{{\rm (5D)}I}D^{\cP\alp}V^J
 +\frac{1}{2J_y}\brkt{\nabla_y^{{\rm (5D)}\cP} V^ID^{\cP\alp}V^J
 -\nabla_y^{{\rm (5D)}\cP}D^{\cP\alp}V^IV^J}}. 
\ee

The field strengths~$\cY_{\rm T\alp}$ and $\cV_{\rmT}$ become 
\bea
 \cY_{\rm T\alp} \eql \SE\cW_{\rm T5\alp}+\SE\nabla_5\Ups_{\rm T\alp} \nonumber\\
 \toa \SE\cW_\alp^T,  \nonumber\\
 \cV_{\rmT} \toa \cV^T \equiv \nabla_y^{{\rm (5D)}\cP}V^T
 -\brkt{J_y\hat{\Sgm}^{{\rm (5D)}T}+\hc}
 +\cO(\Psi_y U^4,(U^4)^2),  \label{limit:cY_T}
\eea
where 
\bea
 \cW_\alp^T \defa \cW_{\rmT 5} = -\frac{1}{4}\bar{D}^2D_\alp V^T, \nonumber\\
 V^T \defa V_{\rmT 5}, \;\;\;\;\;
 \Sgm^{{\rm (5D)}T} \equiv -\Sgm_{\rmT}. 
\eea
Thus, we obtain 
\bea
 \frac{\cX_{\rmT}}{\UE} = \frac{\cV_{\rmT}}{\VE\UE}
 \to \frac{\cV^T |\hat{S}_{\rm E}|^2}{\VE^{\rm (5D)}}, 
\eea
and 
\bea
 &&\frac{\cX_{\rmT}}{\UE}L_{\rm V}^{(3)}
 +\brkt{J_\cP\frac{\cX_{\rmT}}{\UE}L_{\rm V}^{(4)}+\hc} \nonumber\\
 \toa \frac{\cV^T|\hat{S}_{\rm E}|^2}{\VE^{\rm (5D)}}f_{IJ}\left\{
 \frac{4}{|\hat{S}_{\rm E}|^2}\brkt{\frac{1}{J_y}\nabla_y^{{\rm (5D)}\cP}V^I-\hat{\Sgm}^{{\rm (5D)}I}}^\dagger
 \brkt{\frac{1}{J_y}\nabla_y^{{\rm (5D)}\cP}V^J-\hat{\Sgm}^{{\rm (5D)}J}}
 \right. \nonumber\\
 &&\hspace{25mm}\left.
 +\brkt{\frac{2J_y}{\bar{J}_y|\hat{S}_{\rm E}|^2}\hat{\Sgm}^{{\rm (5D)}I}\hat{\Sgm}^{{\rm (5D)}J}+\hc}
 \right\} \nonumber\\
 \eql \frac{2\cV^T f_{IJ}}{\VE^{\rm (5D)}}\cV^I\cV^J, 
\eea
up to $\cO(\Psi_4U^4,(U^4)^2)$, 
where
\be
 \cV^I \equiv \nabla_y^{{\rm (5D)}\cP}V^I-\brkt{J_y\hat{\Sgm}^{{\rm (5D)}I}+\hc}. 
\ee
We have used the limit of $J_S^{(1)}\to 1/\bar{J}_y+\cO((U^4)^2)$. 

As a result, the Lagrangian in the vector-tensor sector becomes 
\bea
 \cL_{\rm VT} \defa \int\dr^4\tht\brc{\brkt{J_\cP L_{\rm V}^{(1)\alp}\hat{\cY}_{\rm T\alp}+\hc}
 +\cV_{\rm T}L_{\rm V}^{(2)}
 +\frac{\cX_{\rmT}}{\UE}L_{\rm V}^{(3)}
 +\brkt{J_\cP\frac{\cX_{\rmT}}{\UE}L_{\rm V}^{(4)}+\hc}} \nonumber\\
 \toa \int\dr^4\tht\;\left[f_{IJ}\left\{-2J_y\hat{\Sgm}^{{\rm (5D)}I}D^{\cP\alp}V^J 
 \right.\right.\nonumber\\
 &&\hspace{20mm}\left.
 +\frac{1}{2J_y}\brkt{\nabla_y^{{\rm (5D)}\cP} V^ID^{\cP\alp}V^J
 -\nabla_y^{{\rm (5D)}\cP}D^{\cP\alp}V^IV^J}\right\}\hat{\cW}_\alp^T  \nonumber\\
 &&\hspace{12mm}\left.
 +f_{IJ}\cV^T\brkt{D^{\cP\alp}V^I\hat{\cW}_\alp^J
 +\frac{1}{2}V^I D^{\cP\alp}\hat{\cW}_\alp^J+\hc} 
 +\frac{2f_{IJ}}{\VE^{\rm (5D)}}\cV^T\cV^I\cV^J\right] \nonumber\\
 \eql \brkt{-\int\dr^2\tht\;C_{\bar{I}\bar{J}\bar{K}}
 \Sgm^{{\rm (5D)}\bar{I}}\cW^{\bar{J}}\cW^{\bar{K}}+\hc} \nonumber\\
 &&+\int\dr^4\tht\;\brc{\frac{C_{\bar{I}\bar{J}\bar{K}}}{3J_y}\brkt{\der_y V^{\bar{I}} D^\alp V^{\bar{J}}
 -\der_y D^\alp V^{\bar{I}}V^{\bar{J}}}\hat{\cW}_\alp^{\bar{K}}+\hc} \nonumber\\
 &&+\int\dr^4\tht\;\frac{2C_{\bar{I}\bar{J}\bar{K}}}{3\VE^{\rm (5D)}}
 \cV^{\bar{I}}\cV^{\bar{J}}\cV^{\bar{K}}, \label{limit:cL_VT}
\eea
up to $\cO(\Psi_4U^4,(U^4)^2)$, 
where the indices~$\bar{I},\bar{J},\bar{K}$ run over $T,0,1,2,\cdots$, 
and the completely symmetric constant tensor~$C_{\bar{I}\bar{J}\bar{K}}$ 
is defined as $C_{IJT}=f_{IJ}$ and the other components are zero. 
This agrees with the 5D result in Ref.~\cite{Sakamura:2012bj} 
at the linearized order in $\Psi_y^\alp$ and $U^4$. 
At the last step in (\ref{limit:cL_VT}), we have used the relation 
\bea
 &&\frac{f_{IJ}}{J_y}\left\{\brkt{\nabla_y^{{\rm (5D)}\cP}V^I D^{\cP\alp}V^J
 -\nabla_y^{{\rm (5D)}\cP}D^{\cP\alp}V^IV^J}\hat{\cW}_{\rmT 5\alp} \right.\nonumber\\
 &&\hspace{10mm}\left.
 +\brkt{\nabla_y^{{\rm (5D)}\cP}V^T D^{\cP\alp}V^I
 -\nabla_y^{{\rm (5D)}\cP}D^{\cP\alp}V^TV^I}\hat{\cW}_\alp^J\right\}+\hc \nonumber\\
 \eql \frac{2f_{IJ}}{J_y}\brkt{\nabla_y^{{\rm (5D)}\cP}V^I D^{\cP\alp}V^T
 -\nabla_y^{{\rm (5D)}\cP}D^{\cP\alp}V^IV^T}\hat{\cW}_\alp^J+\hc, 
\eea
which can be shown in the same way as Appendix~D in Ref.~\cite{Abe:2015yya}.

\subsection{Gravitational sector}
Finally, we consider the gravitational sector. 
Since $\cC_{\rm E}^\mu$ in (\ref{def:cC_E}) becomes 
\be
 \cC_{\rm E}^\mu \to \frac{1}{\SE}\nabla_4 U^\mu 
 = \frac{1}{\SE}\left\{\der_4 U^\mu 
 -\frac{1}{2}\sgm^\mu_{\alp\dalp}\brkt{\bar{D}^{\dalp}\Psi_4^\alp-D^\alp\bar{\Psi}_4^{\dalp}}
 +V_E(\Im\SE^2)\der^\mu U^4\right\}, 
\ee
we find that
\be
 \frac{\bar{\cC}_{\rm E}^\mu\cC_{\rmE\mu}}{2\vev{\VE\UE}} 
 \to \frac{\cC^{\rm (5D)\mu}\cC^{\rm (5D)}_\mu}{2\vev{\VE^{\rm (5D)}}}, 
\ee
where
\be
 \cC^{{\rm (5D)}\mu} \equiv \der_y U^\mu
 -\frac{1}{2}\sgm^\mu_{\alp\dalp}\brkt{\bar{D}^{\dalp}\Psi_y^\alp-D^\alp\bar{\Psi}_y^{\dalp}}
 -\VE^{\rm (5D)}\der^\mu U^4. 
\ee
This agrees with the kinetic term for $U^4$ and $\Psi_y^\alp$ 
in Ref.~\cite{Sakamura:2012bj}. 

Finally, we give a comment on the independence of $\VE^{\rm (5D)}$ defined in (\ref{def:V_E^5D}). 
Notice that $\SE$ disappears in the 5D action, and $\Ups_{\rmT\alp}$ appears 
only through $\cX_{\rmT}$ in $\VE$ after the dimensional reduction. 
(The $\Ups_{\rmT}$-dependence of $\cY_{\rmT\alp}$ disappears as shown in (\ref{limit:cY_T}).)
Thus, although $\VE$ in the 6D SUGRA action is not an independent degree of freedom 
(see (\ref{sol:V_E})), 
$\VE^{\rm (5D)}$ is independent in the 5D SUGRA action. 
Namely, the degrees of freedom of $\SE$ and $\Ups_{\rmT\alp}$ are converted 
into that of $\VE^{\rm (5D)}$.

\section{Summary} \label{summary}
In this paper, we have completed the $\cN=1$ superfield description of 6D SUGRA. 
Specifically, we have clarified the dependence of the action on the $\cN=1$ superfields 
that contain the ``off-diagonal'' components of the sechsbein~$e_m^{\;\;\udl{\nu}}$, $e_\mu^{\;\;\udl{n}}$, 
which were missing in our previous work~\cite{Abe:2015yya}. 
These superfields are necessary for the invariance of the action under 
the full 6D diffeomorphisms and the Lorentz transformations 
in the $\cN=1$ superfield description. 
The corresponding superfields~$U^m$ and $\Psi_m^\alp$ play roles of the gauge fields 
for those transformations. 
Although they do not have zero-modes in many extra-dimensional models, 
they can give significant effects on 4D effective theory 
when they are integrated out, as in the case of 5D SUGRA~\cite{Abe:2008an,Abe:2011rg}. 

Our results are collected in Sec.~\ref{result}. 
The superfields~$U^m$ and $\Psi_m^\alp$ appear in the action 
in a nontrivial manner, but the resultant action is consistent with the 6D diffeomorphisms, 
6D Lorentz transformations and the transformation laws of the component fields. 
Besides, it reduces to the known 5D SUGRA action in Ref.~\cite{Sakamura:2012bj}. 
These properties ensure the reliability of our result. 

In this paper, $\Psi_m^\alp$ are treated at the linearized level. 
This is because we have adopted the linearized 4D SUGRA 
formulation~\cite{Sakamura:2011df,Ferrara:1977mv,Siegel:1978mj} 
to describe the 4D part of the 6D Weyl multiplet. 
In order to treat $\Psi_m^\alp$ at full order, we need to use 
the complete conformal superspace formulation~\cite{Butter:2009cp}, 
which is technically more complicated. 

Our 6D SUGRA description is useful to construct or analyze various setups for the braneworld models 
that contain lower-dimensional branes or the orbifold fixed points. 
Besides, it is also powerful for the systematic derivation of 4D effective action 
that keeps the $\cN=1$ superspace structure. 

We have focused on the case of the Abelian gauge group, for simplicity. 
In order to extend our result to the non-Abelian case, 
we need to include an additional term, 
which is the SUGRA counterpart of (3.9) in Ref.~\cite{Marcus:1983wb}
or (2.23) in Ref.~\cite{ArkaniHamed:2001tb}, 
to ensure the gauge invariance. 

We will discuss these issues in a subsequent paper.

\subsection*{Acknowledgements}
This work was supported in part by JSPS KAKENHI Grant Numbers JP16K05330 (H.A.), 
JP16J06569 (S.A.) and JP25400283 (Y.S.). 

\appendix

\section{$\bdm{\cN=1}$ SUGRA couplings} \label{4DSUGRAcouplings}
In this section, we summarize the result of Ref.~\cite{Sakamura:2011df}, 
and show how to obtain the couplings to the $\cN=1$ SUGRA multiplet. 
This corresponds to the modification of the 4D linearized SUGRA~\cite{Ferrara:1977mv,Siegel:1978mj} 
to make the relation to the superconformal formulation 
in Refs.~\cite{Kaku:1978nz}-\cite{Kugo:1982cu} clearer. 
Before the gauge fixing of the extraneous symmetry, 
the action has the $\cN=1$ superconformal symmetry that consists of the invariance under 
the translation~$\bdm{P}$, SUSY~$\bdm{Q}$, the local Lorentz transformation~$\bdm{M}$, 
the dilatation~$\bdm{D}$, the automorphism~${\rm U(1)}_A$, 
the conformal boost~$\bdm{K}$, and the conformal SUSY~$\bdm{S}$. 
In Ref.~\cite{Sakamura:2011df}, we expressed this formulation in the language of 
the superfields at the linearized order in (the fluctuation part of) the gravitational fields. 
In this appendix, we neglect terms beyond this order, 
and the background spacetime is assumed to be 4D Minkowski spacetime.

\subsection{Definition of superfields}
The independent fields in the Weyl multiplet are 
the vierbein~$e_\mu^{\;\;\udl{\nu}}$, the gravitino~$\psi_{\mu\alp}$, 
the ${\rm U(1)}_A$-gauge field~$A_\mu$, and the $\bdm{D}$-gauge field~$b_\mu$. 
Among them, $b_\mu$ does not play any essential role, and can be set to zero, 
which corresponds to the $\bdm{K}$-gauge fixing.  

The vierbein~$e_\mu^{\;\;\udl{\nu}}$ is divided into the background~$\vev{e_\mu^{\;\;\udl{\nu}}}$ 
and the fluctuation~$\tl{e}_\mu^{\;\;\nu}$ as
\be
 e_\mu^{\;\;\udl{\nu}} = \vev{e_\mu^{\;\;\udl{\rho}}}
 \brkt{\dlt_\rho^{\;\;\nu}+\tl{e}_\rho^{\;\;\nu}},  \label{def:tle}
\ee
where $\vev{e_\mu^{\;\;\udl{\nu}}}=\dlt_\mu^{\;\;\nu}$ by our assumption.\footnote{ 
We need not discriminate the curved indices~$\mu$ from the flat one~$\udl{\mu}$ 
for $\tl{e}_\rho^{\;\;\nu}$ whose Weyl weight is 0. 
}
Then we can form the following real superfield. 
\bea
 U^\mu \eql (\tht\sgm^{\udl{\rho}}\bar{\tht})\vev{e_{\udl{\rho}}^{\;\;\nu}}\tl{e}_\nu^{\;\;\mu}
 +i\bar{\tht}^2\vev{e_{\udl{\rho}}^{\;\;\mu}}
 \brkt{\tht\sgm^{\udl{\nu}}\bar{\sgm}^{\udl{\rho}}\psi_{\udl{\nu}}}
 -i\tht^2\vev{e_{\udl{\rho}}^{\;\;\mu}}
 \brkt{\bar{\tht}\bar{\sgm}^{\udl{\nu}}\sgm^{\udl{\rho}}\bar{\psi}_{\udl{\nu}}} 
 \nonumber\\
 &&+\frac{1}{4}\tht^2\bar{\tht}^2\brkt{3A^\mu-\ep^{\mu\nu\rho\tau}\der_\nu\tl{e}_{\rho\tau}}. 
\eea
We have included $\vev{e_\mu^{\;\;\nu}}$ in the above expression 
in order to make the counting of the Weyl weight clear. 
This superfield has the Weyl weight 0. 

We construct a chiral superfield from a (superconformal) chiral multiplet~$\sbk{\phi,\chi_\alp,F}$ as
\bea
 \Phi \eql \brkt{1+\frac{w}{3}\cE}\brkt{\phi+\tht\chi+\tht^2F}, \nonumber\\
 \cE \defa \tl{e}_\mu^{\;\;\mu}-2i\tht\sgm^{\udl{\mu}}\bar{\psi}_{\udl{\mu}}, 
\eea
where $w$ denotes the Weyl weight (\ie, the $\bdm{D}$ charge) of this multiplet. 

We also construct a real (unconstrained) superfield 
from a real general multiplet~$\sbk{C,\zeta_\alp,\cH,B_\mu,\lmd_\alp,D}$~\footnote{
A complex scalar~$\cH$ is $\frac{1}{2}(H+iK)$ in the notation of Ref.~\cite{Kugo:1982cu}. 
} 
as 
\bea
 V \eql \brc{1+\frac{w}{6}\brkt{\cE+\bar{\cE}}}\left\{
 C+i\tht\zeta-i\bar{\tht}\bar{\zeta}-\tht^2\cH-\bar{\tht}^2\bar{\cH}
 -(\tht\sgm^{\udl{\mu}}\bar{\tht})B'_{\udl{\mu}} \right.\nonumber\\
 &&\hspace{35mm}\left.
 +i\tht^2(\bar{\tht}\bar{\lmd}')-i\bar{\tht}^2(\tht\lmd')+\frac{1}{2}\tht^2\bar{\tht}^2D'\right\}, 
 \nonumber\\
\eea
where
\bea
 B'_\mu \defa B_\mu-\zeta\psi_\mu-\bar{\zeta}\bar{\psi}_\mu-\frac{w}{2}CA_\mu, \nonumber\\
 \lmd'_\alp \defa \lmd_\alp-\frac{i}{2}\brkt{\sgm^\mu\der_\mu\bar{\zeta}}_\alp
 -\brkt{\sgm^\mu\bar{\sgm}^\nu\psi_\mu}_\alp B_\nu
 -\frac{w}{4}\brkt{\sgm^\mu\bar{\zeta}}_\alp A_\mu, \nonumber\\
 D' \defa D-\frac{1}{2}g^{\mu\nu}\der_\mu\der_\nu C+\cdots, 
\eea
and $\sgm_{\alp\dalp}^\mu\equiv\vev{e_{\udl{\nu}}^{\;\;\mu}}\sgm_{\alp\dalp}^{\udl{\nu}}$.

\subsection{Superconformal transformation} \label{N1sf_trf}
With the above definitions of the superfields, 
the (linearized) superconformal transformations are expressed as~\footnote{
We take the metric convention and the definitions of the spinor derivatives of 
Ref.~\cite{Wess:1992cp}, which are different from 
those in Ref.~\cite{Sakamura:2011df}. 
} 
\bea
 \dlt_L U^\mu \eql -\frac{1}{2}\sgm^\mu_{\alp\dalp}
 \brkt{\bar{D}^{\dalp}L^\alp-D^\alp\bar{L}^{\dalp}}, 
 \nonumber\\
 \dlt_L\Phi \eql \brkt{-\frac{1}{4}\bar{D}^2L^\alp D_\alp
 +i\sgm_{\alp\dalp}^\mu\bar{D}^{\dalp}L^\alp\der_\mu-\frac{w}{12}\bar{D}^2D^\alp L_\alp}\Phi 
 \nonumber\\
 \eql -\frac{1}{4}\bar{D}^2\brkt{L^\alp D_\alp\Phi+\frac{w}{3}D^\alp L_\alp \Phi}, 
 \nonumber\\
 \dlt_LV \eql \brkt{-\frac{1}{4}\bar{D}^2L^\alp D_\alp
 +\frac{i}{2}\sgm^\mu_{\alp\dalp}\bar{D}^{\dalp}L^\alp\der_\mu
 -\frac{w}{24}\bar{D}^2D^\alp L_\alp+\hc}V,  \label{dlt_L}
\eea
where the transformation parameter~$L^\alp$ is an unconstrained complex spinor superfield. 
The components of $L^\alp$ denoted as
\bea
 \xi^\mu \defa \left.-\Re\brkt{i\sgm^\mu_{\alp\dalp}\bar{D}^{\dalp}L^\alp}\right|_{\tht=0}, \;\;\;\;\;
 \ep_\alp \equiv \left.-\frac{1}{4}\bar{D}^2L_\alp\right|_{\tht=0}, \nonumber\\
 \lmd_{\mu\nu} \defa \left.-\frac{1}{2}\Re\brc{\brkt{\sgm_{\mu\nu}}_\alp^{\;\;\bt}
 D_\alp\bar{D}^2L^\bt}\right|_{\tht=0}, \;\;\;\;\;
 \vph_D \equiv \left.\Re\brkt{\frac{1}{4}D^\alp\bar{D}^2L_\alp}\right|_{\tht=0}, \nonumber\\
 \vth_A \defa \left.\Im\brkt{-\frac{1}{6}D^\alp\bar{D}^2L_\alp}\right|_{\tht=0}, \;\;\;\;\;
 \eta_\alp \equiv \left.-\frac{1}{32}D^2\bar{D}^2L_\alp\right|_{\tht=0}, 
\eea
represent the transformation parameters for $\bdm{P}$, $\bdm{Q}$, $\bdm{M}$, $\bdm{D}$, 
${\rm U(1)}_A$ and $\bdm{S}$, respectively. 
As we can see from (\ref{dlt_L}), $U^\mu$ transforms nonlinearly, and thus it corresponds to 
the gauge (super)field for the $\dlt_L$-transformation. 
We should also note that this superfield transformation preserves the chirality condition:
$\bar{D}_{\dalp}\Phi=0$.

\subsection{Invariant action} \label{4DInv_action}
For a given global SUSY Lagrangian: 
\bea
 \cL_{\rm matter} \eql \int\dr^4\tht\;\Omg(\Phi,V)
 +\sbk{\int\dr^2\tht\;\brc{W(\Phi)-\frac{1}{4}f(\Phi)\cW^\alp\cW_\alp}+\hc},
\eea
where $\Omg$ is a real function, $W$ and $f$ are holomorphic functions, and 
$\cW_\alp \equiv -\frac{1}{4}\bar{D}^2D_\alp V$, 
we can make it invariant under the $\dlt_L$-transformation 
by inserting $U^\mu$ in the following way.  
\be
 \cL = \int\dr^4\tht\;\brkt{1+\frac{1}{3}E_1}\Omg\brkt{\Phi_U,V}
 +\sbk{\int\dr^2\tht\;\brc{W(\Phi)-\frac{1}{4}f(\Phi)\cW_U^\alp\cW_{U\alp}}+\hc}, 
\ee
where
\bea
 E_1 \defa \frac{1}{4}\bar{\sgm}_\mu^{\dalp\alp}\sbk{D_\alp,\bar{D}_{\dalp}}U^\mu, \;\;\;\;\;
 \bar{\sgm}_\mu^{\dalp\alp} \equiv \vev{e_\mu^{\;\;\udl{\nu}}}\bar{\sgm}_{\udl{\nu}}^{\dalp\alp}, 
 \nonumber\\
 \Phi_U \defa \brkt{1+iU^\mu\der_\mu}\Phi, \nonumber\\
 \cW_{U\alp} \defa -\frac{1}{4}\bar{D}^2\brkt{D_\alp V
 +\frac{1}{4}D_\alp U^\mu\bar{\sgm}_\mu^{\dbt\bt}\sbk{D_\bt,\bar{D}_{\dbt}}V
 -iU^\mu\der_\mu D_\alp V}. 
 \label{def:E_1}
\eea
Here, the operation of $(1+iU^\mu\der_\mu)$ on $\Phi$ is understood as 
the embedding of the chiral multiplet into a general multiplet. 
The modified field strength superfield~$\cW_{U\alp}$ is invariant under the gauge transformation:
\be
 V \to V+\brkt{1+iU^\mu\der_\mu}\Lmd+\brkt{1-iU^\mu\der_\mu}\bar{\Lmd}, 
\ee
where $\Lmd$ is a chiral superfield. 

The kinetic term for $U^\mu$ is given by~\footnote{
The $\bdm{D}$-gauge-fixing condition that leads to the canonically normalized 
Einstein-Hilbert term is given by $\Omg|_{\tht=0}=-3$ in the unit of the Planck mass. 
}
\bea
 \cL_{\rm E}^{N=1} \eql \int\dr^4\tht\;
 \frac{\vev{\Omg}}{3}\brc{\frac{1}{8}U^\mu D^\alp\bar{D}^2D_\alp U_\mu
 +\frac{1}{3}E_1^2-\brkt{\der_\mu U^\mu}^2},  \label{kin_for_U^mu}
\eea
where the Weyl weight of 
$U_\mu=\vev{e_\mu^{\;\;\udl{\rho}}}\vev{e_\nu^{\;\;\udl{\tau}}}\eta_{\udl{\rho}\udl{\tau}}U^\nu$ 
is $-2$. 

Using the above insertion of $U^\mu$, 
the $\cN=1$ (linearized) SUGRA Lagrangian is obtained by choosing 
\bea
 \Omg \eql -3\abs{\Phi_U^{\rm com}}^2e^{-K(\Phi_U,V)/3}, \nonumber\\
 W \eql (\Phi^{\rm comp})^3W_{\rm SUGRA}(\Phi), 
\eea
where $\Phi^{\rm comp}$ is the compensator chiral superfield, $\Phi$ is the physical chiral superfield, 
the real function~$K(\Phi_U,V)$ is the K\"{a}hler potential, 
and the holomorphic function~$W_{\rm SUGRA}(\Phi)$ is the superpotential.

\section{Diffeomorphism of component fields} \label{diffeo:comp}
Under the diffeomorphism, the coordinates and the fields transform as
\bea
 \dlt_\xi x^M \eql \xi^M \nonumber\\
 \dlt_\xi e_M^{\;\;\udl{N}} \eql \xi^L\der_L e_M^{\;\;\udl{N}}+\der_M\xi^L e_L^{\;\;\udl{N}}, 
 \nonumber\\
 \dlt_\xi\phi_i^{\bar{A}} \eql \xi^M\der_M\phi_i^{\bar{A}}, \nonumber\\
 \dlt_\xi A_M^I \eql \xi^N\der_N A_M^I+\der_M\xi^N A_N^I, \nonumber\\
 \dlt_\xi\sgm \eql \xi^M\der_M\sgm, \nonumber\\
 \dlt_\xi B_{MN} \eql \xi^L\der_L B_{MN}+\der_M\xi^L B_{LN}+\der_N\xi^L B_{ML}, \;\;\;\;\; \cdots, 
 \label{diffeo}
\eea
where the transformation parameters~$\xi^M(x)$ are real functions.  
The 6D diffeomorphism~$\dlt_\xi$ can be divided into the 4D part~$\dlt_\xi^{(1)}$ 
with $\xi^\mu$, and the extra-dimensional part~$\dlt_\xi^{(2)}$ with $\xi^m$. 
In this section, we focus on the $\dlt_\xi^{(2)}$-transformations of 
the component fields of the $\cN =1$ superfields.

\subsection{Weyl multiplet}
\ignore{
The sechsbein can be divided into the following submatrices. 
\bea
 e_M^{\;\;\udl{N}} \eql \begin{pmatrix} e_\mu^{\;\;\udl{\nu}} & e_\mu^{\;\;\udl{n}} \\ 
 e_m^{\;\;\udl{\nu}} & e_m^{\;\;\udl{n}} \end{pmatrix} 
 \equiv \begin{pmatrix} e_{(4,4)} & e_{(4,2)} \\ e_{(2,4)} & e_{(2,2)} \end{pmatrix}. 
\eea
The inverse matrix is then expressed as
\bea
 e_{\udl{M}}^{\;\;N} \eql \begin{pmatrix} e_{\udl{\mu}}^{\;\;\nu} & e_{\udl{\mu}}^{\;\;n} \\ 
 e_{\udl{m}}^{\;\;\nu} & e_{\udl{m}}^{\;\;n} \end{pmatrix} 
 = \begin{pmatrix} e_{(4,4)} & e_{(4,2)} \\ e_{(2,4)} & e_{(2,2)} \end{pmatrix}^{-1} \nonumber\\
 \eql \begin{pmatrix} \brkt{e_{(4,4)}-e_{(4,2)}e_{(2,2)}^{-1}e_{(2,4)}}^{-1} &
 -e_{(4,4)}^{-1}e_{(4,2)}\brkt{e_{(2,2)}-e_{(2,4)}e_{(4,4)}^{-1}e_{(4,2)}}^{-1} \\
 -e_{(2,2)}^{-1}e_{(2,4)}\brkt{e_{(4,4)}-e_{(4,2)}e_{(2,2)}^{-1}e_{(2,4)}}^{-1} & 
 \brkt{e_{(2,2)}-e_{(2,4)}e_{(4,4)}^{-1}e_{(4,2)}}^{-1} \end{pmatrix}. \nonumber\\
\eea
By dropping the fluctuation modes in $e_\mu^{\;\;\udl{\nu}}$ around the background~$\vev{e_{(4,4)}}=\id_4$, 
we have
\bea
 e_{\udl{\mu}}^{\;\;\nu} \eql \id_4+e_{(4,2)}e_{(2,2)}^{-1}e_{(2,4)}^{-1}+\cdots, \nonumber\\
 e_{\udl{\mu}}^{\;\;n} \eql -e_{(4,2)}e_{(2,2)}^{-1}+\cdots, \nonumber\\
 e_{\udl{m}}^{\;\;\nu} \eql -e_{(2,2)}^{-1}e_{(2,4)}+\cdots, \nonumber\\
 e_{\udl{m}}^{\;\;n} \eql e_{(2,2)}^{-1}+e_{(2,2)}^{-1}e_{(2,4)}e_{(4,4)}^{-1}e_{(4,2)}e_{(2,2)}^{-1}+\cdots, 
 \label{inverseEs}
\eea
where the ellipses denote terms beyond linear in $e_{(2,4)}$ or $e_{(4,2)}$. 
}
From the second equation in (\ref{diffeo}), 
$E_m\equiv e_m^{\;\;\udl{4}}+ie_m^{\;\;\udl{5}}$ transforms as
\be
 \dlt^{(2)}_\xi E_m = \der_m\xi^n E_n+\xi^n\der_n E_m, 
\ee
which leads to 
\bea
 \dlt^{(2)}_\xi S_E| \eql \xi^m\der_m S_E|+\frac{1}{2}\brkt{\der_4\xi^4-\der_5\xi^5
 +\frac{1}{S_E^2|}\der_4\xi^5-S_E^2|\der_5\xi^4}S_E|, \nonumber\\
 \dlt^{(2)}_\xi\brkt{E_4E_5} \eql \xi^m\der_m\brkt{E_4E_5}
 +\brkt{\der_m\xi^m+\frac{1}{\SE^2|}\der_4\xi^5+\SE^2|\der_5\xi^4}\brkt{E_4E_5}, 
 \label{diff:S_E}
\eea
where $S_E|\equiv\sqrt{E_4/E_5}$. 

Here we impose the constraint:~\footnote{
This constraint preserves the values of $e_m^{\;\;\udl{\mu}}$ under the 4D diffeomorphism, 
but we do not take a gauge in which they are fixed to zero. 
}
\be
 \der_m\xi^{\udl{\mu}} = \der_m\brkt{\xi^N e_N^{\;\;\udl{\mu}}} = 0. 
\ee
Then the ``off-diagonal'' components~$e_m^{\;\;\udl{\mu}}$ transform as 
\bea
 \dlt_\xi e_m^{\;\;\udl{\mu}} \eql \xi^N\der_N e_m^{\;\;\udl{\mu}}
 +\der_m\xi^N e_N^{\;\;\udl{\mu}} \nonumber\\
 \eql \xi^N\der_N e_m^{\;\;\udl{\mu}}-\xi^N\der_m e_N^{\;\;\udl{\mu}}. 
\eea
Namely, its $\dlt_\xi^{(2)}$-transformation is 
\be
 \dlt_\xi^{(2)}e_m^{\;\;\udl{\mu}} = \xi^n\brkt{\der_n e_m^{\;\;\udl{\mu}}-\der_m e_n^{\;\;\udl{\mu}}}. 
 \label{dlt_xi:e_m^umu}
\ee

Since 
\bea
 \dlt_\xi e_{\udl{M}}^{\;\;N} \eql -e_{\udl{M}}^{\;\;L}\brkt{\dlt_\xi e_L^{\;\;\udl{P}}}e_{\udl{P}}^{\;\;N} 
 \nonumber\\
 \eql \xi^P\der_P e_{\udl{M}}^{\;\;N}-e_{\udl{M}}^{\;\;P}\der_P\xi^N
 = \xi^P\der_P e_{\udl{M}}^{\;\;N}-\der_{\udl{M}}\xi^N,  \label{dltxi2:inverseE}
\eea
we obtain
\bea
 \dlt_\xi^{(2)}e_{\udl{\mu}}^{\;\;m} \eql -\der_{\udl{\mu}}\xi^m+\xi^n\der_n e_{\udl{\mu}}^{\;\;m}. 
 \label{diff:e_bmu^m}
\eea

Besides, $e^{(2)}=e_4^{\;\;\udl{4}}e_5^{\;\;\udl{5}}-e_4^{\;\;\udl{5}}e_5^{\;\;\udl{4}}$ transforms as
\be
 \dlt_\xi^{(2)}e^{(2)} = \der_m\brkt{\xi^m e^{(2)}}. \label{dlt_xi:e^2}
\ee
Hence, it follows that
\bea
 \dlt_\xi^{(2)}\tl{V}_{\rmE}| \eql \dlt_\xi^{(2)}\brkt{\frac{e^{(2)2}}{\abs{E_4E_5}}} \nonumber\\
 \eql \frac{2e^{(2)}\dlt_\xi^{(2)}e^{(2)}}{\abs{E_4E_5}}
 -\frac{e^{(2)2}}{\abs{E_4E_5}^3}\Re\brc{\bar{E}_4\bar{E}_5\dlt_\xi^{(2)}\brkt{E_4E_5}} \nonumber\\
 \eql  \xi^m\der_m\tl{V}_{\rmE}|
 +\Re\brkt{\der_m\xi^m-\frac{1}{\SE^2|}\der_4\xi^5-\SE^2|\der_5\xi^4}\tl{V}_{\rmE}|, 
 \label{diff:tlV_E}
\eea
where $\tl{V}_{\rm E}\equiv\VE\UE$.

\subsection{Hypermultiplet}
Combining the third equation in (\ref{diffeo}) with the second equation in (\ref{diff:S_E}), 
we obtain the transformation of 
$H^{\bar{A}}|\equiv (E_4E_5)^{1/4}\phi_2^{\bar{A}}$ as 
\be
 \dlt^{(2)}_\xi H^{\bar{A}}| = \xi^m\der_m H^{\bar{A}}|+\frac{1}{4}
 \brkt{\der_m\xi^m+\frac{1}{\SE^2|}\der_4\xi^5+\SE^2|\der_5\xi^4}H^{\bar{A}}|. 
 \label{diff:H}
\ee

\subsection{Vector multiplet}
Combining the fourth equation in (\ref{diffeo}) with (\ref{dltxi2:inverseE}), 
we can show that
\be
 \dlt^{(2)}_\xi A_{\udl{\mu}}^I = \dlt^{(2)}_\xi\brkt{e_{\udl{\mu}}^{\;\;N} A_N^I} 
 = \xi^n\der_n A_{\udl{\mu}}^I. \label{diff:V}
\ee

As for the extra-dimensional components, we see that
\bea
 \dlt_\xi^{(2)}\Sgm^I| \eql \frac{i}{2}\brkt{-\frac{1}{\SE^2|}A_4^I-A_5^I}\dlt_\xi^{(2)}\SE|
 +\frac{i}{2}\brkt{\frac{1}{\SE|}\dlt_\xi^{(2)}A_4^I-\SE|\dlt_\xi^{(2)}A_5^I} \nonumber\\
 \eql \xi^m\der_m\Sgm^I|
 +\frac{1}{2}\brkt{\der_m\xi^m-\frac{1}{\SE^2|}\der_4\xi^5-\SE^2|\der_5\xi^4}\Sgm^I|, 
 \label{diff:Sgm}
\eea
where $\Sgm^I|=\frac{i}{2}\brkt{\SE^{-1}|A_4^I-\SE|A_5^I}$.

\subsection{Tensor multiplet}
From the last equation in (\ref{diffeo}) and (\ref{dltxi2:inverseE}), we have
\bea
 \dlt_\xi^{(2)} B_{\udl{\mu}\udl{\nu}} \eql \xi^n\der_n B_{\udl{\mu}\udl{\nu}}, \nonumber\\
 \dlt_\xi^{(2)} B_{\udl{\mu}m} \eql \xi^n\der_n B_{\udl{\mu}m}
 +\der_m\xi^n B_{\udl{\mu}n}, \nonumber\\
 \dlt_\xi^{(2)} B_{45} \eql \xi^n\der_n B_{45}+\der_4\xi^4 B_{45}+\der_5\xi^5 B_{45} 
 = \der_n\brkt{\xi^n B_{45}}, \nonumber\\
 \dlt_\xi^{(2)} B_{\udl{4}\udl{5}} \eql \xi^n\der_n B_{\udl{4}\udl{5}}. 
 \label{dlt_xi:Bs}
\eea

\section{Lorentz transformations of component fields} \label{Lorentz:comp}
In this section we see the Lorentz transformations of the component fields 
of the superfields. 

\subsection{Weyl multiplet}
The sechsbein~$e_M^{\;\;\udl{N}}$ transforms as
\be
 \dlt_\lmd e_M^{\;\;\udl{N}} = \lmd^{\udl{N}}_{\;\;\udl{L}}e_M^{\;\;\udl{L}}, 
\ee
where the transformation parameters~$\lmd^{\udl{N}}_{\;\;\udl{L}}$ are real, 
and $\lmd_{\udl{N}\udl{L}}=-\lmd_{\udl{L}\udl{N}}$.\footnote{
The flat indices~$\udl{M},\udl{N},\cdots$ are raised and lowered by 
$\eta^{\udl{M}\udl{N}}$ and $\eta_{\udl{M}\udl{N}}$, respectively. 
} 
In the following, we focus on the transformations by $\lmd^{\udl{\mu}}_{\;\;\udl{n}}$, 
which mix 4D and the extra dimensions.  

First, note that
\bea
 \dlt_\lmd E_m \eql \dlt_\lmd\brkt{e_m^{\;\;\udl{4}}+ie_m^{\;\;\udl{5}}} 
 = \brkt{\lmd^{\udl{4}}_{\;\;\udl{\mu}}+i\lmd^{\udl{5}}_{\;\;\udl{\mu}}}e_m^{\;\;\udl{\mu}} \nonumber\\
 \eql-\brkt{\lmd_{\udl{\mu}\udl{4}}+i\lmd_{\udl{\mu}\udl{5}}}e_m^{\;\;\udl{\mu}}, \nonumber\\
 \dlt_\lmd e^{(2)} \eql \dlt_\lmd\Im\brkt{\bar{E}_4E_5} \nonumber\\
 \eql -\Im\brc{\brkt{\lmd_{\udl{\mu}\udl{4}}-i\lmd_{\udl{\mu}\udl{5}}}
 \brkt{E_5 e_4^{\;\;\udl{\mu}}-E_4 e_5^{\;\;\udl{\mu}}}} \nonumber\\
 \eql -e^{(2)}\Re\brc{{\brkt{\lmd_{\udl{\mu}\udl{4}}-i\lmd_{\udl{\mu}\udl{5}}}
 \frac{\sqrt{E_4E_5}}{ie^{(2)}}\times\brkt{\sqrt{\frac{E_5}{E_4}}e_4^{\;\;\udl{\mu}}
 -\sqrt{\frac{E_4}{E_5}}e_5^{\;\;\udl{\mu}}}}}. 
\eea
Since these are proportional to $e_m^{\;\;\udl{\mu}}$, we can see that
\be
 \dlt_\lmd\sqrt{\frac{E_4}{E_5}} = \cO(e_m^{\;\;\udl{\mu}}), \;\;\;\;\;
 \dlt_\lmd\brkt{\frac{e^{(2)2}}{\abs{E_4E_5}}} 
 = \cO(e_m^{\;\;\udl{\mu}}). 
\ee
These are consistent with the first and the fourth transformations in (\ref{dlt_Ns}) if we choose 
the lowest component of $N$ as zero, $N|=0$. 

In the following, we neglect the ``off-diagonal'' components~$e_m^{\;\;\udl{\nu}}$ and $e_\mu^{\;\;\udl{n}}$ 
in the right-hand sides. 
Then we can see that
\bea
 \dlt_\lmd e_{\udl{\mu}}^{\;\;4} \eql \lmd_{\udl{\mu}}^{\;\;\udl{n}}e_{\udl{n}}^{\;\;4} 
 = \frac{1}{e^{(2)}}\brkt{\lmd_{\udl{\mu}}^{\;\;\udl{4}}e_5^{\;\;\udl{5}}
 -\lmd_{\udl{\mu}}^{\;\;\udl{5}}e_5^{\;\;\udl{4}}} \nonumber\\
 \eql \Re\brc{\frac{1}{e^{(2)}}\brkt{-\lmd_{\udl{\mu}}^{\;\;\udl{5}}-i\lmd_{\udl{\mu}}^{\;\;\udl{4}}}
 \brkt{e_5^{\;\;\udl{4}}+ie_5^{\;\;\udl{5}}}} \nonumber\\
 \eql \Re\brc{\brkt{\lmd_{\udl{\mu}\udl{4}}-i\lmd_{\udl{\mu}\udl{5}}}\frac{\sqrt{E_4E_5}}{ie^{(2)}}
 \times\sqrt{\frac{E_5}{E_4}}}, \nonumber\\
 \dlt_\lmd e_{\udl{\mu}}^{\;\;5} \eql \lmd_{\udl{\mu}}^{\;\;\udl{n}}e_{\udl{n}}^{\;\;5} 
 = \frac{1}{e^{(2)}}\brkt{-\lmd_{\udl{\mu}}^{\;\;\udl{4}}e_4^{\;\;\udl{5}}
 +\lmd_{\udl{\mu}}^{\;\;\udl{5}}e_4^{\;\;\udl{4}}} \nonumber\\
 \eql \Re\brc{\frac{1}{e^{(2)}}\brkt{\lmd_{\udl{\mu}}^{\;\;\udl{5}}+i\lmd_{\udl{\mu}}^{\;\;\udl{4}}}
 \brkt{e_4^{\;\;\udl{4}}+ie_4^{\;\;\udl{5}}}} \nonumber\\
 \eql -\Re\brc{\brkt{\lmd_{\udl{\mu}\udl{4}}-i\lmd_{\udl{\mu}\udl{5}}}
 \frac{\sqrt{E_4E_5}}{ie^{(2)}}\times\sqrt{\frac{E_4}{E_5}}}, 
\eea
which are consistent with the second and the third transformations in (\ref{dlt_Ns}). 
Besides, since 
\bea
 \dlt_\lmd \brkt{\frac{i}{2}e_{m\udl{\mu}}} 
 \eql \frac{i}{2}\brkt{\lmd_{\udl{\mu}}^{\;\;\udl{4}}e_{m\udl{4}}
 +\lmd_{\udl{\mu}}^{\;\;\udl{5}}e_{m\udl{5}}} \nonumber\\
 \eql \frac{i}{2}\Im\brc{\brkt{\lmd_{\udl{\mu}\udl{5}}
 +i\lmd_{\udl{\mu}\udl{4}}}
 \brkt{e_m^{\;\;\udl{4}}+ie_m^{\;\;\udl{5}}}} \nonumber\\
 \eql -\frac{ie^{(2)}}{2}\Im\brc{
 \brkt{\lmd_{\udl{\mu}\udl{4}}-i\lmd_{\udl{\mu}\udl{5}}}
 \frac{E_m}{ie^{(2)}}}, 
\eea
we obtain 
\bea
 \dlt_\lmd\brkt{\frac{i}{2}e_{4\udl{\mu}}} 
 \eql -\frac{ie^{(2)}}{2}\Im\brc{
 \brkt{\lmd_{\udl{\mu}\udl{4}}-i\lmd_{\udl{\mu}\udl{5}}}
 \frac{\sqrt{E_4E_5}}{ie^{(2)}}\times
 \sqrt{\frac{E_4}{E_5}}}, \nonumber\\
 \dlt_\lmd\brkt{\frac{i}{2}e_{5\udl{\mu}}}
 \eql -\frac{ie^{(2)}}{2}\Im\brc{
 \brkt{\lmd_{\udl{\mu}\udl{4}}-i\lmd_{\udl{\mu}\udl{5}}}
 \frac{\sqrt{E_4E_5}}{ie^{(2)}}\times
 \sqrt{\frac{E_5}{E_4}}}, 
\eea
which are consistent with the transformations in the third line of (\ref{dlt_Ns}).

\subsection{Hypermultiplet}
Since 
\bea
 \dlt_\lmd\brc{\brkt{E_4E_5}^{1/4}\phi_2^{\bar{A}}} 
 \eql \frac{\phi_2^{\bar{A}}}{4(E_4E_5)^{3/4}}
 \brkt{E_5\dlt_\lmd E_4+E_4\dlt_\lmd E_5} = \cO(e_m^{\;\;\udl{\mu}}), 
\eea
the transformations in the fourth line of (\ref{dlt_Ns}) are consistent with 
the component transformations. 
(Recall that we have chosen the lowest component of $N$ as zero.)

\subsection{Vector multiplet}
We can also see that the last two transformations in (\ref{dlt_Ns}) are consistent with 
the $\dlt_\lmd$-transformations of the component fields because 
\bea
 \dlt_\lmd A_{\udl{\mu}}^I \eql \lmd_{\udl{\mu}}^{\;\;\udl{n}}A_{\udl{n}}^I 
 = \lmd_{\udl{\mu}}^{\;\;\udl{4}}\brkt{e_{\udl{4}}^{\;\;4}A_4^I+e_{\udl{4}}^{\;\;5}A_5^I}
 +\lmd_{\udl{\mu}}^{\;\;\udl{5}}\brkt{e_{\udl{5}}^{\;\;4}A_4^I+e_{\udl{5}}^{\;\;5}A_5^I} \nonumber\\
 \eql \lmd_{\udl{\mu}\udl{4}}\frac{1}{e^{(2)}}\brkt{e_5^{\;\;\udl{5}}A_4^I-e_4^{\;\;\udl{5}}A_5^I}
 +\lmd_{\udl{\mu}\udl{5}}\frac{1}{e^{(2)}}\brkt{-e_5^{\;\;\udl{4}}A_4^I+e_4^{\;\;\udl{4}}A_5^I} \nonumber\\
 \eql \frac{1}{e^{(2)}}\brc{\lmd_{\udl{\mu}\udl{4}}\Im\brkt{E_5A_4^I-E_4A_5^I}
 -\lmd_{\udl{\mu}\udl{5}}\Re\brkt{E_5A_4^I-E_4A_5^I}} \nonumber\\
 \eql \Re\brc{\frac{1}{ie^{(2)}}\brkt{\lmd_{\udl{\mu}\udl{4}}-i\lmd_{\udl{\mu}\udl{5}}}
 \brkt{E_5A_4^I-E_4A_5^I}} \nonumber\\
 \eql 2\Im\brc{\brkt{\lmd_{\udl{\mu}\udl{4}}-i\lmd_{\udl{\mu}\udl{5}}}
 \frac{\sqrt{E_4E_5}}{ie^{(2)}}\times
 \frac{i}{2}\brkt{\sqrt{\frac{E_5}{E_4}}A_4^I-\sqrt{\frac{E_4}{E_5}}A_5^I}}, 
\eea
\bea
 \dlt_\lmd\brc{\frac{i}{2}\brkt{\sqrt{\frac{E_5}{E_4}}A_4^I-\sqrt{\frac{E_4}{E_5}}A_5^I}} 
 \eql \dlt_N\brc{-\frac{e^{(2)}}{2\sqrt{E_4E_5}}\brkt{A_{\udl{4}}^I+iA_{\udl{5}}^I}} \nonumber\\
 \eql -\frac{e^{(2)}}{2\sqrt{E_4E_5}}\brkt{
 \lmd_{\udl{4}}^{\;\;\udl{\mu}}+i\lmd_{\udl{5}}^{\;\;\udl{\mu}}}A_{\udl{\mu}}^I \nonumber\\
 \eql \frac{1}{2}\frac{e^{(2)}}{\sqrt{E_4E_5}}\brkt{\lmd_{\udl{\mu}\udl{4}}+i\lmd_{\udl{\mu}\udl{5}}}
 A^{I\udl{\mu}} \nonumber\\
 \eql -\frac{i}{2}\frac{e^{(2)2}}{\abs{E_4E_5}}
 \brc{-\brkt{\lmd_{\udl{\mu}\udl{4}}+i\lmd_{\udl{\mu}\udl{5}}}
 \frac{\sqrt{\bar{E}_4\bar{E}_5}}{ie^{(2)}}}A^{I\udl{\mu}} \nonumber\\
 \eql -\frac{i}{2}\times\frac{e^{(2)2}}{\abs{E_4E_5}}\times
 \brc{\brkt{\lmd_{\udl{\mu}\udl{4}}-i\lmd_{\udl{\mu}\udl{5}}}
 \frac{\sqrt{E_4E_5}}{ie^{(2)}}}^*\times A^{I\udl{\mu}}, \nonumber\\
\eea
and 
\bea
 \left.-\frac{i}{8}\bar{D}^2\brkt{\tl{V}_{\rm E}D^\alp\bar{N}D_\alp V^I}\right| 
 \eql -\frac{i}{2}\times\tl{V}_{\rm E}|\times\Lmd_{\udl{\mu}}\times A^{I\udl{\mu}}. 
\eea
where $\Lmd_{\udl{\mu}}$ denotes the $\tht\bar{\tht}$-component of $N$, \ie, 
\be
 \Lmd_{\udl{\mu}} \equiv \brkt{\lmd_{\udl{\mu}\udl{4}}-i\lmd_{\udl{\mu}\udl{5}}}
 \frac{\sqrt{E_4E_5}}{ie^{(2)}}. 
\ee


\end{document}